%% file: main.tex
\documentclass[aps,prl,twocolumn,showpacs,superscriptaddress,nofootinbib,preprintnumbers]{revtex4-2}%
\usepackage{graphicx}
\usepackage{tabularx}
\usepackage{bm}
\usepackage{latexsym}
\usepackage{epsf}
\usepackage{rotating}
\usepackage{epsfig,graphics,rotate,color}
\usepackage{wrapfig}
\usepackage{amssymb}
\usepackage{amsmath}
\usepackage{amsfonts}
\usepackage[normalem]{ulem}
\usepackage{color}
\usepackage{style}
\usepackage{siunitx}
\usepackage{placeins}

\begin{document}

\title{First Measurement of Missing Energy due to Nuclear Effects in Monoenergetic Neutrino Charged-Current Interactions}

\include{authorlist}

\begin{abstract}
We present the first measurement of the missing energy due to nuclear effects in monoenergetic, muon neutrino charged-current interactions on carbon, originating from $K^+ \rightarrow \mu^+ \nu_\mu$ decay at rest ($E_{\nu_\mu}=235.5$~MeV), performed with the J-PARC Sterile Neutrino Search at the J-PARC Spallation Neutron Source liquid scintillator based experiment.
Toward characterizing the neutrino interaction, ostensibly $\nu_\mu n \rightarrow \mu^-  p$ or
    $\nu_\mu$$^{12}\mathrm{C}$ $\rightarrow \mu^-$$^{12}\mathrm{N}$, we define the \textit{missing} energy as the energy
    transferred to the nucleus ($\omega$) minus the kinetic energy of the outgoing proton(s), $E_{m} \equiv
    \omega-\sum T_p$, and relate this to visible energy in the detector,
    $E_{m}=E_{\nu_\mu}~(235.5~\mathrm{MeV})-m_\mu~(105.7~\mathrm{MeV}) + [m_n-m_p~(1.3~\mathrm{MeV})] - E_{\mathrm{vis}}$.
    The missing energy, which is naively expected to be zero in the absence of nuclear effects
    (e.g. nucleon separation energy, Fermi momenta, and final-state interactions), is
    uniquely sensitive to many aspects of the interaction, and has previously been inaccessible with
    neutrinos.
The shape-only, differential cross section measurement reported, based on a
$(77\pm3)$\% pure double-coincidence kaon decay-at-rest signal (621~total events), provides
detailed insight into neutrino-nucleus interactions, allowing even the nuclear
orbital shell of the struck nucleon to be inferred.
The measurement provides an important benchmark for models and event generators
at hundreds of MeV neutrino energies, characterized by the difficult-to-model
transition region between neutrino-nucleus and neutrino-nucleon scattering, and
relevant for applications in nuclear physics, neutrino oscillation measurements,
and Type-II supernova studies.
\end{abstract}
\maketitle

\section{Introduction}
Models of neutrino-nucleus interactions are a crucial part of all accelerator-based neutrino physics programs
performing oscillation measurements.
Such models, usually implemented via a neutrino event generator, are used to form multiple aspects
of each measurement, including expected signal and background rates, observable event topologies,
and detector efficiency, bias and resolution.
As we enter an era of precision measurements, uncertainties in neutrino
interaction physics need to be controlled at the percent level or better for next generation
experiments to achieve their physics goals~\cite{NuSTECWhitePaper}.
In recognition of this challenge, there is an ongoing global program of dedicated neutrino
interaction measurements, across many energies and with an array of nuclear
targets~\cite{NuScatteringReview}.
Monoenergetic (235.5~MeV) muon neutrinos from charged kaon decay at rest (KDAR; $K^+ \rightarrow \mu^+
\nu_\mu$, with a branching ratio of 63.6\%~\cite{pdg}) represent a unique and important part of this
program, in particular for informing interaction models in the hundreds of MeV region~\cite{Spitz:2014hwa}.
At these energies, the utilization of such models is wide-ranging, with relevance across multiple current and future neutrino oscillation experiments~\cite{MiniBooNE:2008paa, MicroBooNE:2016pwy, MicroBooNE:2015bmn,
T2K:2011qtm, Cao:2014bea, Hyper-KamiokandeProto-:2015xww, ESSnuSB:2021azq} and understanding Type-II supernovae and the
neutrino-induced nucleosynthesis processes inside~\cite{Nikolakopoulos:2020alk}.

KDAR neutrinos are an important tool for studying neutrino interactions and the nuclear response because their energy is
known, unlike all
other relevant sources of neutrinos above the $\nu_\mu$ charged-current (CC) threshold. These neutrinos are also particularly interesting because their characteristic energy transfer range ($0<\omega<120$~MeV) lies in the
difficult-to-model transition region between neutrino-on-nucleus and neutrino-on-nucleon scattering, in which the interaction evolves from inducing collective nuclear excitations among multiple
nucleons to quasielastic scattering off of individual nucleons. In addition, along with neutrino interaction and nuclear physics, KDAR neutrinos have been proposed as a signature of dark matter annihilation in the Sun~\cite{Rott:2015nma,DUNE:2021gbm} and as a source for neutrino oscillation searches at short and long baseline~\cite{Spitz:2012gp,Axani:2015dha,Grant:2015jva,Harnik:2019iwv}.

Despite this widespread applicability, there exists only one measurement of KDAR neutrinos to date,
based on a 3.9$\sigma$ observation of the process and coarsely presented in terms of shape-only muon
kinetic energy $\frac{1}{\sigma}\frac{d\sigma}{dT_\mu}$~\cite{MiniBooNE:2018dus}.
However, several detailed cross-section calculations have been performed for KDAR neutrino
interactions on carbon and argon~\cite{Martini:2009uj,Martini:2011wp,Nikolakopoulos:2020alk,
KDAR-RPA-2017} and a number of neutrino event generators can be used to simulate these events.
Unfortunately, and as exemplified in Ref.~\cite{Nikolakopoulos:2020alk}, the predictions vary
significantly;
depending on the generator or model, the total KDAR $\nu_\mu$-carbon cross sections range from
$0.9-1.8\times 10^{-39}~\mathrm{cm}^2/\mathrm{neutron}$ and agreement among the anticipated
kinematic distributions [e.g. muon energy ($E_\mu$) and angle ($\theta_\mu$)]  is similarly weak. In general, the large disagreement among predictions underscores the importance of measurements of this process for both fully elucidating the neutrino-nucleus interaction and precisely studying oscillations involving hundreds of MeV neutrinos.

The J-PARC Sterile Neutrino Search at the J-PARC Spallation Neutron Source (JSNS$^2$) experiment
employs a 48~ton liquid scintillator detector [96 inner photomultiplier tubes (PMTs) and 24 veto PMTs] at 24~m from the
Materials and Life Science spallation neutron source originating from 3~GeV protons focused onto a
mercury target, currently operating at 950~kW.
Proton interactions with the target, surrounded mainly by concrete and iron shielding, readily
produce pions, muons, and kaons to create the primarily decay-at-rest neutrino source--about
97.8\% of neutrino parent $K^+$ come to rest before decaying.
In particular, JSNS$^2$ uses the muon decay-at-rest component ($\mu^+ \rightarrow e^+ \overline{\nu}_\mu \nu_e$) to search for possible signatures of short-baseline oscillations near $\Delta m^2 \sim 1$\,eV$^{2}$ ($\overline{\nu}_\mu \rightarrow \overline{\nu}_e$; $\overline{\nu}_e p \rightarrow e^+ n$).
Most relevant for this paper, the protons produce KDAR neutrinos at the rate of
0.0038~KDAR~$\nu_\mu$ per proton on target (POT), according to a detailed, high statistics, \texttt{Geant4}~\cite{Geant4}
simulation of the target and shielding geometry. Notably, however, kaon production at the source is
highly uncertain, with the \texttt{MARS} simulation package~\cite{osti_1282121} predicting nearly a factor of
2 higher rate.

The JSNS$^2$ experiment~\cite{JSNS2:2013jdh,Ajimura:2017fld} began taking data in July 2020 and has gathered $4.9\times10^{22}$~POT as of June 2024. The detector is described in detail in
Ref.~\cite{JSNS2:2021hyk}.
JSNS$^2$ is sensitive to the scintillation light produced in neutrino
interactions inside of the 17~ton cylindrical target volume, composed of linear alkylbenzene (LAB), 3~g/L PPO, 15~mg/L bis-MSB, and loaded at 0.1\% concentration with
gadolinium.
The surrounding 31~ton buffer volume is composed of a similar liquid
scintillator, but without gadolinium and with 30~mg/L bis-MSB;
the outer buffer volume is instrumented to veto events originating outside the target volume.
For the run period analyzed here, an additional solvent di-isopropylnaphthalene (DIN) was added to the target volume at the 8\% level to enhance pulse shape discrimination ability, for differentiating fast-neutron-like and electronlike signals.

For KDAR $\nu_\mu$ CC events, we associate the visible energy, after accounting for quenching, light propagation, and other detector effects, to the
primary interaction products via $E_{\mathrm{vis}}\approx T_\mu+\sum T_p$, where $T_{\mu}$ and $T_{p}$ represent the kinetic energy associated with the produced muon and proton(s) respectively, noting that multiple protons
can be produced via final-state interactions (FSI).
Small ($\lesssim$1\,MeV) contributions due to nuclear deexcitation gammas,
interactions of possible FSI-induced fast neutrons, and the nuclear recoil
energy render this simplistic association between ``visible energy'' and muon and proton kinematics imperfect; these effects can slightly alter the intrinsically inclusive detector observable $E_{\mathrm{vis}}$ and our measurement includes, and does not subtract, these contributions since they are difficult to model. This is discussed more below.
Naively, assuming that the incoming KDAR neutrino interacts quasielastically
with a neutron to produce a muon and a proton, $\nu_\mu n \rightarrow \mu^- p$,
one might assume that the calorimetric energy is simply
$E_{\mathrm{vis,naive}}=E_{\nu_\mu} (235.5~\mathrm{MeV})-m_\mu~(105.7~\mathrm{MeV}) + [m_n-m_p]~(1.3~\mathrm{MeV})  = 131.2~\mathrm{MeV}$.
However, nuclear physics effects, in particular nucleon separation energy and
Fermi momenta, as well as Pauli blocking and FSI, among others, significantly
modify this expectation.
As shown below, predictions for this modification among generators and models vary considerably.

In this article, we report the first measurement of this missing energy due to
nuclear effects using neutrinos, quantified as
$E_{m} \equiv \omega-\sum T_p =
E_{\nu_\mu}~(235.5~\mathrm{MeV})-m_\mu~(105.7~\mathrm{MeV}) + [m_n-m_p~(1.3~\mathrm{MeV})] - E_{\mathrm{vis}}$.
This is analogous to the missing energy variable commonly used to describe
quasielastic events in electron scattering [e.g., Ref.~\cite{Makins:1994mm} with $^{12}$C$(e,e'p)$],
noting that our definition of missing energy contains contributions from both the interaction cross section and nuclear effects, including FSI.

\section{Reconstruction, Simulation, and Analysis}
This analysis is based on $1.3\times10^{22}$~POT taken from January--June 2021.
During this data taking period, the beam and detector performed reliably and were stable as determined by a variety of metrics, including light yield based on muon-decay Michel electrons, regular \textit{in situ} LED-based calibrations, neutron capture on gadolinium signals, beam timing, singles flash rate, neutrino candidates per proton on target, and slow control monitoring.
Most notably, we find that the effective light yield varied by as much as 2.6\% over the course of the run, which we correct for and assign
a systematic uncertainty to.

Our analysis seeks to identify and reconstruct KDAR $\nu_\mu$ CC
events, characterized by a ``prompt" scintillation light flash originating from
the muon and proton(s) ($\sim$20--150~MeV), followed by a ``delayed" Michel
electron signal (0--53~MeV; $\tau_{\mu}=2.0$~$\mu$s~\cite{MuonLifetimePaper}), to form a differential
cross-section measurement in terms of $E_{m}$,
\begin{equation}
(\frac{d\sigma}{dE_{m}})_i \sim \frac{\sum_j M_{ij}(d_j-b_j)}{\epsilon_i},
\end{equation}
where the indices $i$ and $j$ respectively represent true and reconstructed $E_{m}$ (based on
the prompt flash detector observable, $E_{\mathrm{prompt}}$), $d_j$ is the measured data
distribution passing selection, $b_j$ is the predicted background passing
selection, $M_{ij}$ is the unfolding matrix to transform from reconstructed to
true $E_{m}$, and $\epsilon_i$ is the reconstruction efficiency.
Note that we report a \textit{shape-only}
differential cross section, $\frac{1}{\sigma}(\frac{d\sigma}{dE_{m}})_i$.
The choice to ignore normalization is based on the highly uncertain kaon
production at the source previously mentioned.

Towards forming this measurement, we utilize a PMT-charge-based
maximum-likelihood reconstruction technique to relate the measured PMT pulses to
the position and energy of the event.
This reconstruction procedure relies on likelihood functions representing the
probabilities to measure detector observables as a function of the underlying
event quantities and determined using a detailed detector simulation~\cite{jrjordan_thesis, Lee:2024pwg}.
The simulation is largely based on \texttt{Geant4} with light propagation model inputs, including PMT geometry and response, Birks' constant for electrons and muons, attenuation
length, and others, from the Daya Bay and RENO
experiments~\cite{DayaBay:2012fng,RENO:2012mkc} and a number of \textit{in situ}
calibrations, including those mentioned above and with a $^{252}$Cf source
deployed across a range of vertical positions in the detector~\cite{Lee:2024pwg}.

\begin{figure}
\centering
\includegraphics[width=9.3cm]{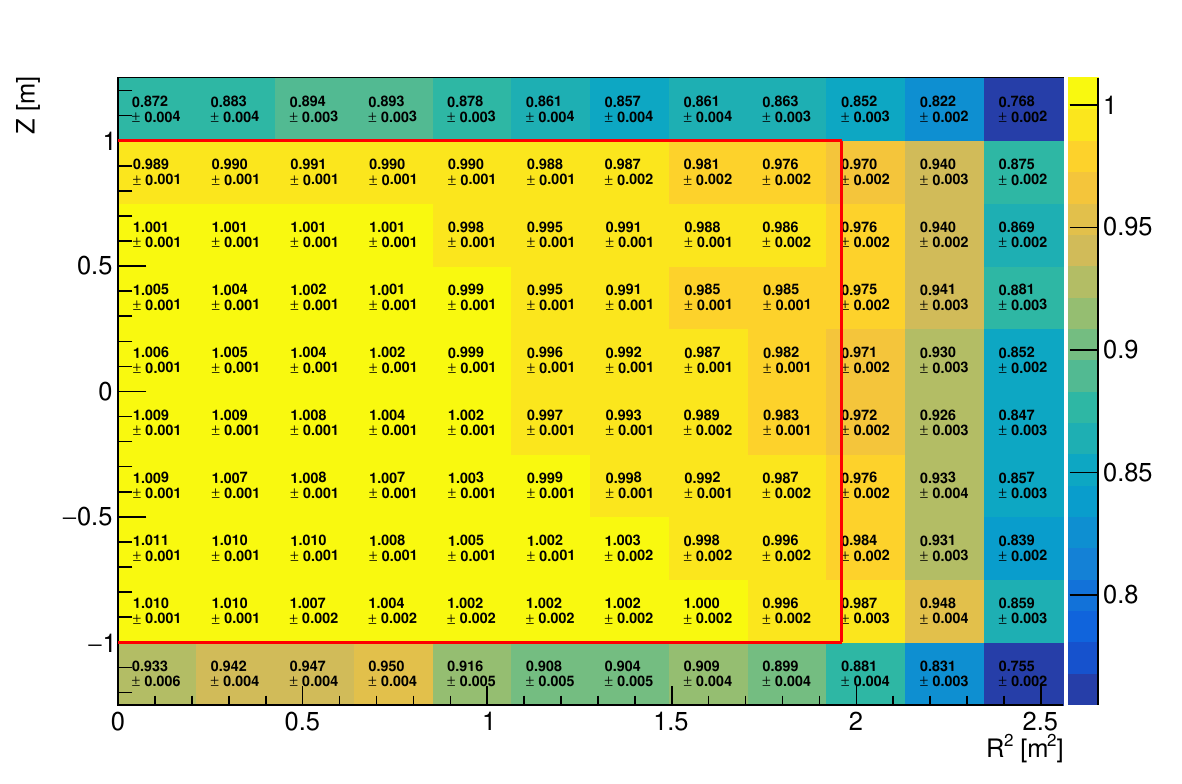}
\includegraphics[width=9.3cm]{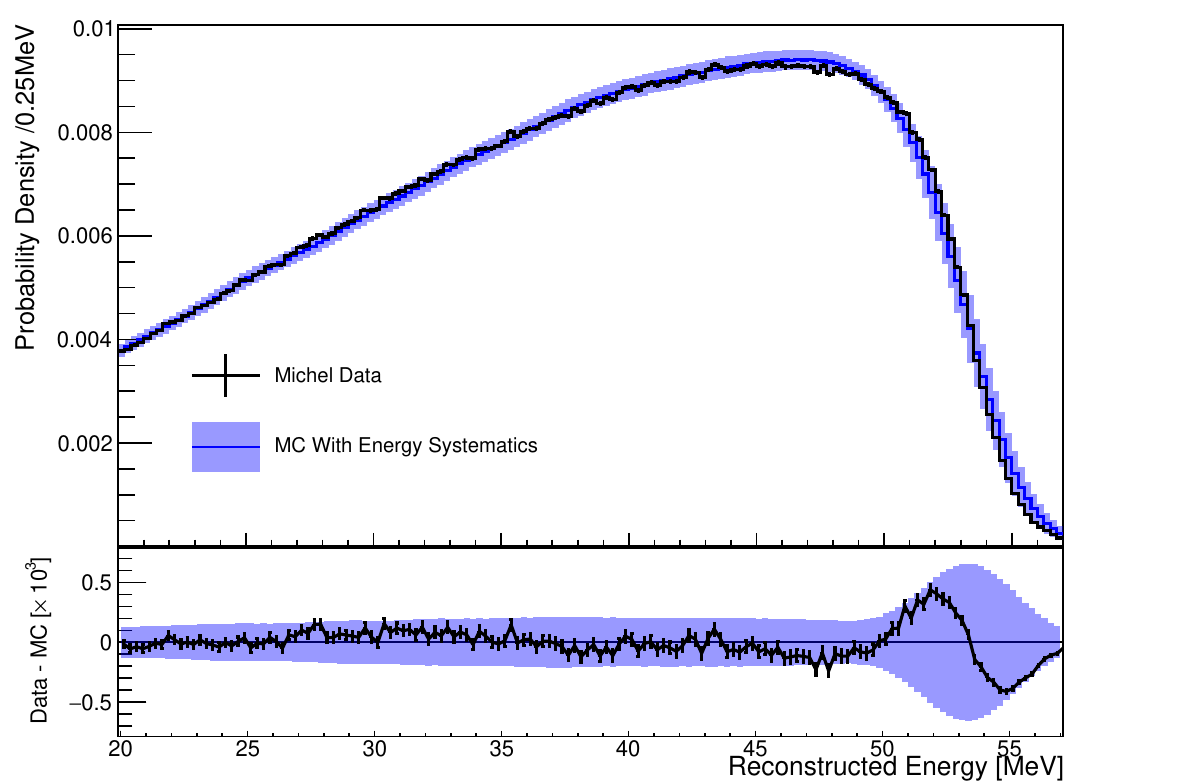}
\caption{Top: observed Michel electron energy scale as a function of position relative to the middle-most position bin; the value is indicated both by the histogram color and bin text. The red box indicates the fiducial volume used for energy calibration.
The coordinates Z and R are the cylindrical height and radius relative to the center of the detector volume, respectively.
Bottom: observed (black) and MC simulated (blue) Michel electron energy
spectrum across the fiducial volume. The blue error band corresponds to the energy reconstruction
systematic uncertainties on the energy scale and resolution applied to the simulated events.}\label{figure_michel_syst}
\end{figure}

In particular, cosmic-induced Michel electrons provide an excellent, high
statistics calibration source throughout the detector.
We compare Monte Carlo (MC) simulated Michel electron events to detected events by fitting each
with an analytic function for the Michel spectrum, convolved with an energy
resolution function, with fit parameters for resolution and energy scale.
The relative energy resolution is modeled as $\frac{\sigma{(E)}}{E} =
\sqrt{\sigma_{\mathrm{ep}}^{2}\frac{E_{\mathrm{ep}}}{E} + C^{2}}$, where
$\sigma_{\mathrm{ep}}$ is the resolution at the endpoint energy ($E_{\mathrm{ep}}$) and is a free parameter
in the fit and $C$ represents the constant minimum possible energy resolution,
estimated to be $\sim$2.5\%.
Notably, while the observed endpoint energy resolution is (3.79$\pm$0.08)\%, the MC simulated resolution
is determined to be (2.78$\pm$0.06)\%; we apply an additional energy smearing to MC events
to account for this discrepancy.
We also compare the energy scale as a function of position throughout
the inner volume between MC simulated Michel electron events and observed
events, and correct the MC events' reconstructed energy to compensate
for the observed discrepancy between the two.
Figure~\ref{figure_michel_syst} shows the energy scale discrepancy from the nominal value throughout the detector.

For each correction to the energy scale and/or resolution we estimate its uncertainty, and account for these as systematics on the energy reconstruction.
We furthermore account for a potential energy scale nonlinearity by comparing
the scale observed for neutron capture on gadolinium events ($E\approx8$\,MeV) and
for Michel electron events ($E_{\mathrm{ep}}\approx53$\,MeV).
The observed energy scale nonlinearity matches the predicted
nonlinearity from simulation to within 0.2\%.
We scale the MC simulated energy response, at energies between the neutron capture energy and Michel endpoint energy, to compensate for this nonlinearity uncertainty, then use the difference between that transformed MC simulation's response at higher energies and the standard MC simulation's response, to estimate the corresponding impact on the KDAR analysis.
At these higher energies we expect electronics saturation and nonlinearity in the PMTs’ charge response to become increasingly prevalent.
These effects are modeled within the MC simulation and reconstruction software to mitigate their contribution to the energy scale non-linearity.
We see good agreement between the MC simulation and the observed data for the rate of saturated PMT signals in the KDAR dataset.
The final energy scale and energy resolution systematic uncertainties (at the Michel endpoint energy) are 0.68\% and 0.31\%, respectively.
Figure~\ref{figure_michel_syst} also shows the observed Michel electron energy
spectrum and the corresponding simulation with systematic uncertainties
applied.

For simulating neutrino events, we utilize the
 \texttt{NuWro} (v21, with $^{12}$C grid spectral function)~\cite{Golan:2012wx,Golan:2012rfa} and \texttt{GiBUU} (v2021p1)~\cite{Lalakulich:2011eh,Buss:2011mx} event generators combined with the detector simulation to find the detection efficiency ($\epsilon_i$) and unfolding matrix ($M_{ij}$) described above.
 The Michel electron calibration results and systematic uncertainties are used to constrain the MC simulation energy response. For all the simulations we include both the quasielastic scattering process as well as the meson exchange current interaction, noting that collective excitations within the nucleus are not modeled by the simulation software.
The two software packages also do not include nuclear deexcitation photons.
The \texttt{NucDeEx} simulation~\cite{NucDeEx, TALYS} predicts that 73\% of KDAR events will produce no deexcitation gammas, with a mean expected energy contribution from deexcitation gammas of 0.75\,MeV.

\begin{figure}[t]
\begin{centering}
\includegraphics[width=9.3cm]{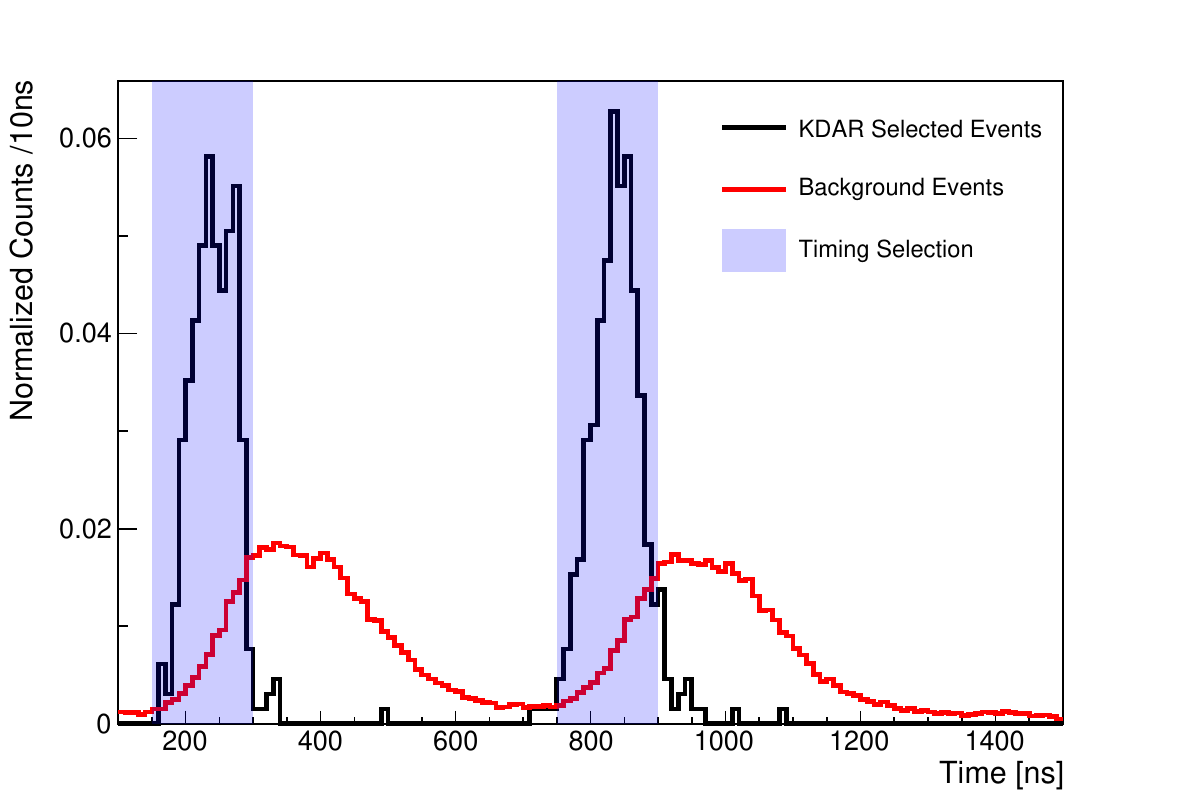}
\vspace{-.7cm}
\caption{Measured distributions of signallike (black) and backgroundlike
    (red) events, based on whether or not they pass selection criteria other than
    timing.
The two pulses corresponding to the selection's proton-on-target arrival windows
    are shown.
The background distribution shows the prompt, signallike ($20<E_{\mathrm{prompt}}<150$~MeV)
    events that are rejected by the nontiming selection criteria--most of these are due to beam fast neutrons.
    }
\label{figure_kdar_signal_timing}
\end{centering}
\end{figure}

Signal-induced prompt events are searched for in a timing window corresponding
to the beam-on-target arrival plus time of flight.
The beam strikes the target with two adjacent pulses, each reasonably consistent with a Landau-Gaussian convolution ($1\sigma \approx40$\,ns), separated by 600\,ns, at 25\,Hz.
We open two corresponding 150\,ns windows around the expected KDAR arrival times
for selecting events, also in consideration of the $K^+$ lifetime
($\tau_{K^+}=12.4$\,ns).
Figure~\ref{figure_kdar_signal_timing} shows the measured beam timing
distributions for signal prompt events passing all KDAR selection criteria other
than timing, further discussed below, and background events that are rejected by
other selection criteria.
The delayed signal due to the muon-decay Michel electron is searched for in a 10~$\mu$s
timing window after the prompt beam pulse.
In addition to timing selection of the two signal flashes, we also require
candidate events to have (1) reconstructed energies $20<E_{\mathrm{prompt}}<150$~MeV and
$20<E_{\mathrm{delayed}}<60$~MeV, with the lower limits set by background mitigation and
signal efficiency considerations and the upper limits set by kinematic thresholds; (2) a primary reconstructed ``vertex," or more
accurately ``mean gamma emission point (MGEP)", in a fiducial volume, described with cylindrical coordinates relative to the center of the detector, of $\mathrm{R}<1.4$~m and
$-1.0<\mathrm{Z}<0.5$~m, where the boundaries of the target volume are $0<\mathrm{R}<1.6$~m and
$-1.25<\mathrm{Z}<1.25$~m; (3) a reconstructed distance between the primary flash and
delayed flash MGEP ($\Delta_{\mathrm{MGEP}}$) of $<30$~cm; and (4) not be associated
with veto activity, noting that a 10~$\mu$s dead time is imposed after a detected
cosmic muon~\cite{hyoungku_thesis}.
In the dataset, 621 prompt and delayed event pairs meet all of the selection criteria.

The selection criteria are imposed to mitigate background contributions from
accidental coincidences, cosmics, beam-induced fast neutrons, and beam-induced non-KDAR neutrino and antineutrino events.
Background events from accidental coincidences, primarily from uncorrelated
cosmic and/or beam-related activity, are characterized by purposefully producing
prompt-delayed pairs with events from different beam spills, treating them as
though they were from the same beam spill.
True correlated signals do not appear across beam signals, so any apparent
correlation produced with this scheme will be accidental.
The mispaired events are subjected to the KDAR selection criteria to estimate
the rate and spectral shape of the accidental background.
By this method the accidental background rate is estimated to be $<$1.14 events
in the KDAR dataset.
Backgrounds from true correlated events, produced by cosmogenic activity (or any
other ambient source), are characterized using data taken when the beam was
off. Applying the KDAR event selection to these data produces the estimated
rate and spectral shape for the cosmogenic background. The cosmogenic background
rate within the analyzed dataset is estimated to be $<$6.57 events.

Beam-produced fast neutrons are a potential background source since they can interact within the target volume to produce a pion and subsequent decay muon and Michel electron, thus potentially faking
the prompt and delayed signature of a KDAR event. However, beam neutrons enter the detector relatively late compared to
KDAR events because most are nonrelativistic,
so we can constrain the fast-neutron rate by looking for an excess of events
with late timing; we find that this background is negligible and consistent with zero.

The most significant background is from beam non-KDAR $\nu_\mu$ and $\overline{\nu}_\mu$ that are
sufficiently energetic to undergo the same CC interaction as the KDAR neutrinos, the dominant source of which is from positively charged pion and kaon decay in flight ($\pi^{+}
\rightarrow \mu^+ \nu_\mu$, $K^+ \rightarrow \mu^+
\nu_\mu$). 
Notably, the timing, $E_{\mathrm{prompt}}$, MGEP, and $\Delta_{\mathrm{MGEP}}$ characteristics of the observed background in a kinematically disallowed KDAR sideband region ($\Delta_{\mathrm{MGEP}}>30$~cm and $E_{\mathrm{prompt}}>150$~MeV) are highly consistent with non-KDAR $\nu_\mu$/$\overline{\nu}_\mu$ CC events. We estimate this background using a neutrino flux provided by a \texttt{Geant4} simulation of the beam-target interaction and the
surrounding target geometry.
Uncertainty on the neutrino flux is estimated by performing the simulation using
both the \texttt{QGSP-BERT}~\cite{Wright:2015xia} and the \texttt{QGSP-BIC}~\cite{binary_cascade} physics lists within \texttt{Geant4}--the difference between the two simulation results is used as a 1$\sigma$ flux shape uncertainty.
The background neutrino interactions are simulated using both the \texttt{NuWro} and \texttt{GiBUU} event generators, once again taking the difference between their predictions as a systematic uncertainty.

\begin{figure}[t]
\begin{centering}
\includegraphics[width=9.3cm]{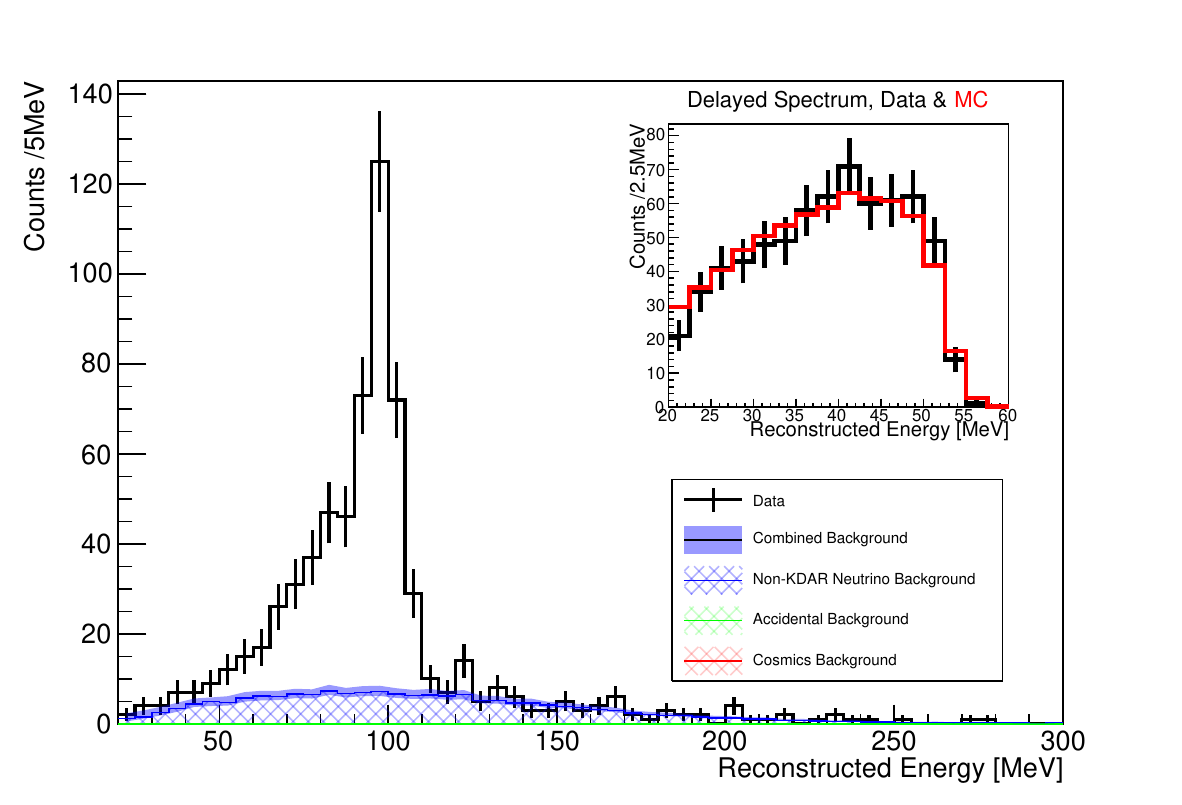}
\vspace{-.7cm}
\caption{After selection, the reconstructed KDAR prompt energy ($E_{\mathrm{prompt}}$) spectrum  including the remaining background components.
The solid blue error band represents the 1$\sigma$ uncertainty on the summed background rate in each energy bin.
The inset shows the reconstructed energy spectrum of delayed Michel signals among KDAR candidates in black, and the red shows the scaled MC simulated KDAR delayed spectrum.}
\label{figure_measured_signal_and_michel}
\end{centering}
\end{figure}

The final estimates for the  signal and background rates come from a simultaneous fit to
the KDAR visible energy spectrum, including sideband constraints, and further
forcing the signal rate to zero in the kinematically disallowed region,
$E_{\mathrm{prompt}}>150$\,MeV.
Markov Chain Monte Carlo sampling is used to evaluate the plausible range of values for the signal and background rates in each energy bin.
The signal rate and the three background event rates in each energy bin are treated as independent parameters, resulting in 26 signal rate parameters and 168 background event rate parameters.
The decay-in-flight background simulations are used as inputs to the fit to constrain the corresponding background shape.
The estimated background shape is allowed to vary smoothly between the four simulation estimates with the observed data providing additional constraint.

The fit estimates the KDAR signal selection to be $(77\pm3)$\% pure with
$144.4^{+21.3}_{-21.1}$ background counts in the signal region.
The total background contribution is estimated to be 99\% from non-KDAR neutrinos.
The fit prefers a non-KDAR neutrino background spectrum more similar to that predicted by the \texttt{NuWro} event generator because \texttt{GiBUU} predicts significantly higher relative rates at low energies, which is inconsistent with the data.
The fit does not have sufficient sensitivity to distinguish meaningfully between the two flux models considered.
Figure~\ref{figure_measured_signal_and_michel} shows the KDAR-like selected events and the results of the background estimate, and the inset shows the spectrum for KDAR candidate delayed Michel electron events
compared to the MC prediction.

The background estimate and uncertainties are used to estimate the underlying KDAR spectrum, which is then unfolded from reconstructed energy, $E_{\mathrm{prompt}}$, to
$E_{m}$.
The detector response, including detection efficiency, is estimated using MC
simulated KDAR events with \texttt{NuWro} and \texttt{GiBUU} and taking their
difference as a source of systematic uncertainty. Most notably, since proton light yield is quenched by the detector's scintillator relative to that of muons, the choice of generator affects the unfolding matrix because the two generators predict different typical distributions of energy splitting between the proton(s) and muons for events with the same $E_{\mathrm{vis}}$.
We use iterative Bayes (D'Agostini)
unfolding~\cite{iterative_unfolding1,dagostini_unfolding, pyunfold} with a $p >
0.9$ between the refolded and the observed distribution used as the stopping
criteria, resulting in four iterations for \texttt{NuWro} and six for
\texttt{GiBUU}.
The smaller number of unfolding iterations indicates that the \texttt{NuWro} KDAR prediction agrees better with our data; we therefore use the \texttt{NuWro} unfolded spectrum and the corresponding errors as our central value result; the \texttt{GiBUU} unfolded result is used to represent a source of systematic uncertainty (described further below).
The \texttt{GiBUU} unfolded result has errors that are, on average, 23\% larger than  the \texttt{NuWro} unfolded result.
Notably, unfolding provides an imperfect estimate characterized by a bias-variance trade-off; because the \texttt{NuWro} and \texttt{GiBUU} predictions are so dissimilar the generator uncertainty contains both the uncertainty due to the difference in estimated detector response and uncertainty due to unfolding bias.

We also consider a systematic uncertainty based on the scintillator's proton Birks'
constant, which sets how much of the proton's energy deposit gets turned into scintillation light. For the detector simulation we use a proton Birks' constant value of 0.097\,mm/MeV, from
Ref.~\cite{vonKrosigk:2013sa} and based on a consistent LAB scintillator admixture, but without DIN; we assign a systematic uncertainty to this value based
on measurements with pure DIN~\cite{din_birks}.
We conservatively take the difference between the two Birks' constant values to be the
magnitude of the systematic error, $kB_{\mathrm{proton}} = 0.097\pm0.011$\,mm/MeV.

Each of the systematic uncertainties are propagated by modifying the MC simulated KDAR
events used to produce the unfolding matrix and those used for the non-KDAR neutrino background estimate.
The systematically varied response matrix and background spectral shape
estimates are used for the background subtraction and unfolding.
The difference between the central value unfolded result and the systematically
varied result is taken as the corresponding systematic uncertainty on the
differential cross section.
For the energy systematic, this process is repeated many times, randomly varying the energy scale and resolution according to their systematic uncertainty.
The Birks' constant uncertainty is evaluated at the estimated $\pm$1$\sigma$ points. The neutrino generator systematic uncertainty is assigned by taking the difference between the \texttt{NuWro} and \texttt{GiBUU} unfolded results as a two-sided, symmetric uncertainty.

\begin{figure}
\begin{centering}
\includegraphics[width=9.3cm]{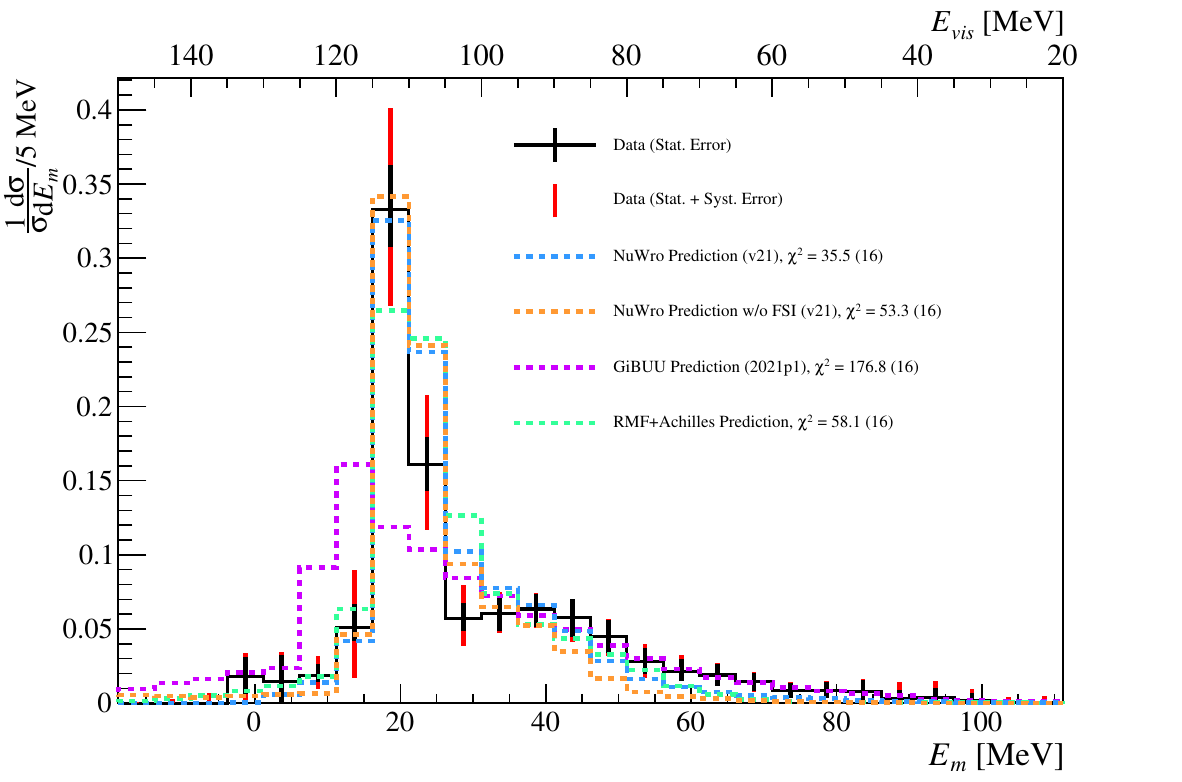}
\vspace{-.7cm}
\caption{The KDAR $\nu_\mu$ CC missing energy, $E_m$, shape-only differential cross-section measurement compared to several neutrino event generator and model predictions. The top x axis provides the corresponding $E_{\mathrm{vis}}$ for each $E_{m}$ value.
}\label{missing_energy_xsec}
\end{centering}
\end{figure}

Figure~\ref{missing_energy_xsec} shows the observed differential cross section in terms of
$E_{m}$ and $E_{\mathrm{vis}}$ for the KDAR $\nu_\mu$ CC interaction on carbon with statistical and systematic error
bars compared to predictions from \texttt{NuWro} (with carbon spectral function), \texttt{GiBUU}, and a relativistic mean field
(RMF) prediction~\cite{rmf_simulation,rmf_spectral_function, PhysRevC.105.025502} using the \texttt{ACHILLES} event generator to simulate FSI~\cite{achilles, achilles_inc, alexis_nikolakopoulos_pc}.
While statistical errors dominate in most of the measurement bins, uncertainties associated with the energy scale and resolution, proton Birks' constant, and generator uncertainty (affecting the unfolding matrix) contribute significantly in the peak region.
This measurement faces limitations from the nuclear modeling available, noting also that the models are \textit{a posteriori} found to be inconsistent with the data;
we hope in the future that a wider variety of sophisticated neutrino-nuclear models become available so that the measurement can incorporate a more complete set of model uncertainties.

The $\chi^2=\sum_{i,j}(d_i-p_i)V_{ij}^{-1}(d_j-p_j)$ comparisons between the data ($d$) and predictions ($p$), using the full covariance matrix ($V_{ij}$) are reported in the figure as well.
For simplicity, we consider the higher statistics region of $E_m$=5--85\,MeV across 16 bins ($i/j$) when forming the $\chi^2$.
As can be seen, the generator and model predictions disagree with each other substantially and none of them is able to adequately match the data well.
While it is difficult to isolate a single cause for this discrepancy across all measurement bins considered, it is much more likely that multiple uncertain nuclear effects are contributing, including FSI, the momentum and separation energy distributions of the nucleons, short-range correlations among nucleons, and even aspects of the weak interaction itself.
Notably, however, the data do favor a significant FSI contribution, with a large $\chi^2$ improvement after turning on this effect within the \texttt{NuWro} event generator, which best matches the data overall.
Included as part of this publication is Supplementary Material that provides the cross-section result as well as covariance matrices for each quantified source of uncertainty.
Also provided is the background subtracted measurement before unfolding, the simulation job cards used for the \texttt{NuWro} and \texttt{GiBUU} simulations, and example analysis software.

The analysis allows for events with missing energy less than zero; the observation prefers a small amount of events in that region but is consistent with zero.
We also note that the primary peak occurs at $E_{m}\approx 18$\,MeV and the secondary peak occurs at $E_{m} \approx 38$\,MeV, consistent with the estimated removal energy for, respectively, p-shell and s-shell nucleons~\cite{PhysRevD.109.072006, ankowski_sf}. In general, this measurement serves as a unique and first-of-its-kind benchmark, simultaneously sensitive to multiple aspects of the event (including what nuclear shell the neutrino interacted with), towards elucidating low energy neutrino-nucleus interactions.

\section{Conclusion}
We have presented the first measurement of missing energy for monoenergetic $\nu_\mu$ CC interactions based on a $(77\pm3)$\% pure sample of 621~total collected KDAR candidate events.
The shape-only differential cross section extracted disagrees significantly with the generator and model predictions considered.
This result serves as a known-neutrino-energy, standard candle for improving our understanding of difficult to model low-energy neutrino interactions at the transition between on-nucleon and on-nucleus scattering.
In the future, JSNS$^2$ and JSNS$^2$-II~\cite{Ajimura:2020qni} are poised to
follow up on this initial measurement with vastly improved statistics, better
control over systematic uncertainties, in particular the proton's Birks'
constant and energy scale and resolution, and the ability to distinguish KDAR
$\nu_\mu$ CC events with and without neutrons.

\section{Acknowledgements}
We deeply thank the J-PARC for their support,
especially for the MLF and the accelerator groups to provide
an excellent environment for this experiment.
We acknowledge the support of the Ministry of Education, Culture, Sports, Science, and Technology (MEXT) and the JSPS grants-in-aid: 16H06344, 16H03967 23K13133,
24K17074 and 20H05624, Japan. The work is also supported by the National Research Foundation of Korea (NRF): 2016R1A5A1004684, 17K1A3A7A09015973, 017K1A3A7A09016426, 2019R1A2C3004955, 2016R1D1A3B02010606, 017R1A2B4011200, 2018R1D1A1B07050425, 2020K1A3A7A09080133, 020K1A3A7A09080114, 2020R1I1A3066835, 2021R1A2C1013661, 2021R1A6A1A03043957, NRF-2021R1C1C2003615, 2022R1A5A1030700, RS-2023-00212787 and RS-2024-00416839. Our work has also been supported by a fund from the BK21 of the NRF. The University of Michigan gratefully acknowledges the support of the Heising-Simons Foundation. This work conducted at Brookhaven National Laboratory was supported by the U.S. Department of Energy under Contract DE-AC02-98CH10886. The work of the University of Sussex is supported by the Royal Society grant no. IESnR3n170385. We also thank the Daya Bay collaboration for providing the Gd-LS, the RENO collaboration for providing the LS and PMTs, CIEMAT for providing the splitters, Drexel University for providing the FEE circuits and Tokyo Inst. Tech for providing FADC boards.
We would also like to thank Alexis Nikolakopoulos for providing the \texttt{RMF+ACHILLES} simulation sample to us.
\clearpage
\bibliography{main}
\end{document}

%% file: authorlist.tex
\newcommand{\affilkek}{\affiliation{High Energy Accelerator Research Organization (KEK), 1-1 Oho, Tsukuba, Ibaraki, 305-0801, Japan}}
\newcommand{\affilsussex}{\affiliation{Department of Physics and Astronomy, University of Sussex, Falmer, Brighton, BN1 9RH, U.K.}}
\newcommand{\affilasrc}{\affiliation{Advanced Science Research Center, JAEA, 2-4 Shirakata, Tokai-mura, Naka-gun, Ibaraki 319-1195, Japan}}
\newcommand{\affiljaea}{\affiliation{J-PARC Center, JAEA, 2-4 Shirakata, Tokai-mura, Naka-gun, Ibaraki 319-1195, Japan}}
\newcommand{\affilkitasato}{\affiliation{Department of Physics, Kitasato University, 1 Chome-15-1 Kitazato, Minami Ward, Sagamihara, Kanagawa, 252-0373, Japan}}
\newcommand{\affilkyoto}{\affiliation{Department of Physics, Kyoto Sangyo University, Motoyama, Kamigamo, Kita-Ku, Kyoto-City, 603-8555, Japan}}
\newcommand{\affilosaka}{\affiliation{Research Center for Nuclear Physics, Osaka University, 10-1 Mihogaoka, Ibaraki, Osaka, 567-0047, Japan}}
\newcommand{\affiltohoku}{\affiliation{Research Center for Neutrino Science, Tohoku University, 6-3 Azaaoba, Aramaki, Aoba-ku, Sendai 980-8578, Japan}}
\newcommand{\affiltsukuba}{\affiliation{Faculty of Pure and Applied Sciences, University of Tsukuba,\\ Tennodai 1-1-1, Tsukuba, Ibaraki, 305-8571, Japan}}
\newcommand{\affilchonbuk}{\affiliation{Division of Science Education, Jeonbuk National University, 567 Baekje-daero, Deokjin-gu, Jeonju-si, Jeollabuk-do, 54896, Korea}}
\newcommand{\affilchonnam}{\affiliation{Department of Physics, Chonnam National University, 77, Yongbong-ro, Buk-gu, Gwangju, 61186, Korea}}
\newcommand{\affildongshin}{\affiliation{Laboratory for High Energy Physics, Dongshin University, 67, Dongshindae-gil, Naju-si, Jeollanam-do, 58245, Korea}}
\newcommand{\affilgist}{\affiliation{Department of Physics and Photon Science, Gwangju Institute of Science and Technology, 123 Cheomdangwagi-ro, Buk-gu, Gwangju, 61005, Korea}}
\newcommand{\affilkyungpook}{\affiliation{Department of Physics, Kyungpook National University, 80 Daehak-ro, Buk-gu, Daegu, 41566, Korea}}
\newcommand{\affilsungkyunkwan}{\affiliation{Department of Physics, Sungkyunkwan University, 2066, Seobu-ro, Jangan-gu, Suwon-si, Gyeonggi-do, 16419, Korea}}
\newcommand{\affilkyunghee}{\affiliation{Department of Physics, Kyung Hee University, 26, Kyungheedae-ro, Dongdaemun-gu, Seoul 02447, Korea}}
\newcommand{\affilsnu}{\affiliation{School of Liberal Arts, Seoul National University of Science and Technology, 232 Gongneung-ro, Nowon-gu, Seoul, 139-743, Korea}}
\newcommand{\affilseoyeong}{\affiliation{Department of Fire Safety, Seoyeong University, 1 Seogang-ro, Buk-gu, Gwangju, 61268, Korea}}
\newcommand{\affilsoongsil}{\affiliation{Department of Physics and OMEG Institute, Soongsil University, 369 Sangdo-ro, Dongjak-gu, Seoul, 06978, Korea}}
\newcommand{\affilsunyatsen}{\affiliation{School of Physics, Sun Yat-sen (Zhongshan) University, Haizhu District, Guangzhou, 510275, China}}
\newcommand{\affilbnl}{\affiliation{Brookhaven National Laboratory, Upton, NY, 11973-5000, USA}}
\newcommand{\affilutah}{\affiliation{Department of Physics \& Astronomy, The University of Utah, UT, 84112, USA}}
\newcommand{\affilalabama}{\affiliation{University of Alabama, Tuscaloosa, AL, 35487, USA}}
\newcommand{\affilmichigan}{\affiliation{University of Michigan, Ann Arbor, MI, 48109, USA}}

\author{E.\,Marzec} \affilmichigan
\author{S.\,Ajimura} \affilosaka
\author{A.\,Antonakis} \affilmichigan
\author{M.\,Botran} \affilmichigan
\author{M.K.\,Cheoun} \affilsoongsil
\author{J.H.\,Choi} \affildongshin
\author{J.W.\,Choi} \affilchonnam
\author{J.Y.\,Choi} \affilchonnam
\author{T.\,Dodo} \affiltohoku
\author{H.\,Furuta}\affiltohoku
\author{J.H.\,Goh} \affilkyunghee
\author{K.\,Haga} \affiljaea
\author{M.\,Harada} \affiljaea
\author{S.\,Hasegawa} \affilasrc \affiljaea 
\author{Y.\,Hino} \thanks{Now at Okayama University} \affiltohoku 
\author{T.\,Hiraiwa} \thanks{Now at RIKEN} \affilosaka
\author{W.\,Hwang} \affilkyunghee
\author{T.\,Iida} \affiltsukuba
\author{E.\,Iwai} \thanks{Now at RIKEN} \affilmichigan
\author{S.\,Iwata} \thanks{Now at Tokyo Metropolitan College of Industrial Technology (Tokyo Metro. Col. of Indus. Tech.)} \affilkitasato
\author{H.I.\,Jang} \affilseoyeong
\author{J.S.\,Jang} \affilgist
\author{M.C.\,Jang} \affilchonnam
\author{H.K.\,Jeon} \affilsungkyunkwan
\author{S.H.\,Jeon} \thanks{Now at Kyungpook National University} \affilsungkyunkwan
\author{K.K.\,Joo} \affilchonnam
\author{D.E.\,Jung} \affilsungkyunkwan
\author{S.K.\,Kang} \affilsnu
\author{Y.\,Kasugai} \affiljaea
\author{T.\,Kawasaki} \affilkitasato
\author{E.J.\,Kim} \affilchonbuk
\author{J.Y.\,Kim} \affilchonnam
\author{E.M.\,Kim} \affilchonnam
\author{S.Y.\,Kim} \affilchonnam
\author{W.\,Kim} \affilkyungpook
\author{S.B.\,Kim } \affilsunyatsen
\author{H.\,Kinoshita} \affiljaea
\author{T.\,Konno} \affilkitasato
\author{K.\,Kuwata}\affiltohoku
\author{D.H.\,Lee} \affilkek
\author{S.\,Lee} \affilkyunghee
\author{I.T.\,Lim} \affilchonnam
\author{C.\,Little} \affilmichigan
\author{T.\,Maruyama} \affilkek
\author{S.\,Masuda} \affiljaea
\author{S.\,Meigo} \affiljaea
\author{S.\,Monjushiro} \affilkek
\author{D.H.\,Moon} \affilchonnam
\author{T.\,Nakano} \affilosaka
\author{M.\,Niiyama} \affilkyoto
\author{K.\,Nishikawa} \thanks{Deceased}  \affilkek
\author{M.\,Noumachi} \affilosaka
\author{M.Y.\,Pac} \affildongshin
\author{B.J.\,Park} \affilkyungpook
\author{H.W.\,Park} \affilchonnam
\author{J.B.\,Park} \affilsoongsil
\author{J.S.\,Park} \affilkyungpook
\author{J.S.\,Park} \affilchonnam
\author{R.G.\,Park} \affilchonnam
\author{S.J.M.\,Peeters} \affilsussex
\author{G.\,Roellinghoff} \affilsungkyunkwan
\author{C.\,Rott} \affilutah
\author{J.W.\,Ryu} \affilkyungpook
\author{K.\,Sakai} \affiljaea
\author{S.\,Sakamoto} \affiljaea
\author{T.\,Shima} \affilosaka
\author{C.D.\,Shin} \affilkek
\author{J.\,Spitz} \email[Corresponding author ]{spitzj@umich.edu} \affilmichigan
\author{I.\,Stancu} \affilalabama
\author{F.\,Suekane}\affiltohoku
\author{Y.\,Sugaya} \affilosaka
\author{K.\,Suzuya} \affiljaea
\author{M.\,Taira} \affilkek
\author{Y.\,Takeuchi}\affiltsukuba
\author{W.\,Wang} \affilsunyatsen
\author{J.\,Waterfield} \affilsussex
\author{W.\,Wei} \affilsunyatsen
\author{R.\,White} \affilsussex
\author{Y.\,Yamaguchi} \affiljaea
\author{M.\,Yeh} \affilbnl
\author{I.S.\,Yeo} \affildongshin
\author{C.\,Yoo} \affilkyunghee
\author{I.\,Yu} \affilsungkyunkwan
\author{A.\,Zohaib} \affilchonnam
\collaboration{JSNS$^{2}$ Collaboration} \noaffiliation

%% file: main.bbl
\begin{thebibliography}{56}%
\makeatletter
\providecommand \@ifxundefined [1]{%
 \@ifx{#1\undefined}
}%
\providecommand \@ifnum [1]{%
 \ifnum #1\expandafter \@firstoftwo
 \else \expandafter \@secondoftwo
 \fi
}%
\providecommand \@ifx [1]{%
 \ifx #1\expandafter \@firstoftwo
 \else \expandafter \@secondoftwo
 \fi
}%
\providecommand \natexlab [1]{#1}%
\providecommand \enquote  [1]{``#1''}%
\providecommand \bibnamefont  [1]{#1}%
\providecommand \bibfnamefont [1]{#1}%
\providecommand \citenamefont [1]{#1}%
\providecommand \href@noop [0]{\@secondoftwo}%
\providecommand \href [0]{\begingroup \@sanitize@url \@href}%
\providecommand \@href[1]{\@@startlink{#1}\@@href}%
\providecommand \@@href[1]{\endgroup#1\@@endlink}%
\providecommand \@sanitize@url [0]{\catcode `\\12\catcode `\$12\catcode `\&12\catcode `\#12\catcode `\^12\catcode `\_12\catcode `\%12\relax}%
\providecommand \@@startlink[1]{}%
\providecommand \@@endlink[0]{}%
\providecommand \url  [0]{\begingroup\@sanitize@url \@url }%
\providecommand \@url [1]{\endgroup\@href {#1}{\urlprefix }}%
\providecommand \urlprefix  [0]{URL }%
\providecommand \Eprint [0]{\href }%
\providecommand \doibase [0]{https://doi.org/}%
\providecommand \selectlanguage [0]{\@gobble}%
\providecommand \bibinfo  [0]{\@secondoftwo}%
\providecommand \bibfield  [0]{\@secondoftwo}%
\providecommand \translation [1]{[#1]}%
\providecommand \BibitemOpen [0]{}%
\providecommand \bibitemStop [0]{}%
\providecommand \bibitemNoStop [0]{.\EOS\space}%
\providecommand \EOS [0]{\spacefactor3000\relax}%
\providecommand \BibitemShut  [1]{\csname bibitem#1\endcsname}%
\let\auto@bib@innerbib\@empty
\bibitem [{\citenamefont {Alvarez-Ruso}\ \emph {et~al.}(2018)\citenamefont {Alvarez-Ruso} \emph {et~al.}}]{NuSTECWhitePaper}%
  \BibitemOpen
  \bibfield  {author} {\bibinfo {author} {\bibfnamefont {L.}~\bibnamefont {Alvarez-Ruso}} \emph {et~al.} (\bibinfo {collaboration} {NuSTEC}),\ }\bibfield  {title} {\bibinfo {title} {{NuSTEC White Paper: Status and challenges of neutrino\textendash{}nucleus scattering}},\ }\href {https://doi.org/10.1016/j.ppnp.2018.01.006} {\bibfield  {journal} {\bibinfo  {journal} {Prog. Part. Nucl. Phys.}\ }\textbf {\bibinfo {volume} {100}},\ \bibinfo {pages} {1} (\bibinfo {year} {2018})},\ \Eprint {https://arxiv.org/abs/1706.03621} {arXiv:1706.03621 [hep-ph]} \BibitemShut {NoStop}%
\bibitem [{\citenamefont {Mahn}\ \emph {et~al.}(2018)\citenamefont {Mahn}, \citenamefont {Marshall},\ and\ \citenamefont {Wilkinson}}]{NuScatteringReview}%
  \BibitemOpen
  \bibfield  {author} {\bibinfo {author} {\bibfnamefont {K.}~\bibnamefont {Mahn}}, \bibinfo {author} {\bibfnamefont {C.}~\bibnamefont {Marshall}},\ and\ \bibinfo {author} {\bibfnamefont {C.}~\bibnamefont {Wilkinson}},\ }\bibfield  {title} {\bibinfo {title} {{Progress in Measurements of 0.1\textendash{}10 GeV Neutrino\textendash{}Nucleus Scattering and Anticipated Results from Future Experiments}},\ }\href {https://doi.org/10.1146/annurev-nucl-101917-020930} {\bibfield  {journal} {\bibinfo  {journal} {Ann. Rev. Nucl. Part. Sci.}\ }\textbf {\bibinfo {volume} {68}},\ \bibinfo {pages} {105} (\bibinfo {year} {2018})},\ \Eprint {https://arxiv.org/abs/1803.08848} {arXiv:1803.08848 [hep-ex]} \BibitemShut {NoStop}%
\bibitem [{\citenamefont {Zyla}\ \emph {et~al.}(2020)\citenamefont {Zyla} \emph {et~al.}}]{pdg}%
  \BibitemOpen
  \bibfield  {author} {\bibinfo {author} {\bibfnamefont {P.~A.}\ \bibnamefont {Zyla}} \emph {et~al.} (\bibinfo {collaboration} {Particle Data Group}),\ }\bibfield  {title} {\bibinfo {title} {{Review of Particle Physics}},\ }\href {https://doi.org/10.1093/ptep/ptaa104} {\bibfield  {journal} {\bibinfo  {journal} {PTEP}\ }\textbf {\bibinfo {volume} {2020}},\ \bibinfo {pages} {083C01} (\bibinfo {year} {2020})}\BibitemShut {NoStop}%
\bibitem [{\citenamefont {Spitz}(2014)}]{Spitz:2014hwa}%
  \BibitemOpen
  \bibfield  {author} {\bibinfo {author} {\bibfnamefont {J.}~\bibnamefont {Spitz}},\ }\bibfield  {title} {\bibinfo {title} {{Cross Section Measurements with Monoenergetic Muon Neutrinos}},\ }\href {https://doi.org/10.1103/PhysRevD.89.073007} {\bibfield  {journal} {\bibinfo  {journal} {Phys. Rev. D}\ }\textbf {\bibinfo {volume} {89}},\ \bibinfo {pages} {073007} (\bibinfo {year} {2014})},\ \Eprint {https://arxiv.org/abs/1402.2284} {arXiv:1402.2284 [physics.ins-det]} \BibitemShut {NoStop}%
\bibitem [{\citenamefont {Aguilar-Arevalo}\ \emph {et~al.}(2009)\citenamefont {Aguilar-Arevalo} \emph {et~al.}}]{MiniBooNE:2008paa}%
  \BibitemOpen
  \bibfield  {author} {\bibinfo {author} {\bibfnamefont {A.~A.}\ \bibnamefont {Aguilar-Arevalo}} \emph {et~al.} (\bibinfo {collaboration} {MiniBooNE}),\ }\bibfield  {title} {\bibinfo {title} {{The MiniBooNE Detector}},\ }\href {https://doi.org/10.1016/j.nima.2008.10.028} {\bibfield  {journal} {\bibinfo  {journal} {Nucl. Instrum. Meth. A}\ }\textbf {\bibinfo {volume} {599}},\ \bibinfo {pages} {28} (\bibinfo {year} {2009})},\ \Eprint {https://arxiv.org/abs/0806.4201} {arXiv:0806.4201 [hep-ex]} \BibitemShut {NoStop}%
\bibitem [{\citenamefont {Acciarri}\ \emph {et~al.}(2017)\citenamefont {Acciarri} \emph {et~al.}}]{MicroBooNE:2016pwy}%
  \BibitemOpen
  \bibfield  {author} {\bibinfo {author} {\bibfnamefont {R.}~\bibnamefont {Acciarri}} \emph {et~al.} (\bibinfo {collaboration} {MicroBooNE}),\ }\bibfield  {title} {\bibinfo {title} {{Design and Construction of the MicroBooNE Detector}},\ }\href {https://doi.org/10.1088/1748-0221/12/02/P02017} {\bibfield  {journal} {\bibinfo  {journal} {JINST}\ }\textbf {\bibinfo {volume} {12}}\bibfield  {number} {\bibinfo  {number} { (02)},\ \bibinfo {pages} {P02017}},\ }\Eprint {https://arxiv.org/abs/1612.05824} {arXiv:1612.05824 [physics.ins-det]} \BibitemShut {NoStop}%
\bibitem [{\citenamefont {Antonello}\ \emph {et~al.}(2015)\citenamefont {Antonello} \emph {et~al.}}]{MicroBooNE:2015bmn}%
  \BibitemOpen
  \bibfield  {author} {\bibinfo {author} {\bibfnamefont {M.}~\bibnamefont {Antonello}} \emph {et~al.} (\bibinfo {collaboration} {MicroBooNE, LAr1-ND, ICARUS-WA104}),\ }\bibfield  {title} {\bibinfo {title} {{A Proposal for a Three Detector Short-Baseline Neutrino Oscillation Program in the Fermilab Booster Neutrino Beam}},\ }\href@noop {} {\  (\bibinfo {year} {2015})},\ \Eprint {https://arxiv.org/abs/1503.01520} {arXiv:1503.01520 [physics.ins-det]} \BibitemShut {NoStop}%
\bibitem [{\citenamefont {Abe}\ \emph {et~al.}(2011)\citenamefont {Abe} \emph {et~al.}}]{T2K:2011qtm}%
  \BibitemOpen
  \bibfield  {author} {\bibinfo {author} {\bibfnamefont {K.}~\bibnamefont {Abe}} \emph {et~al.} (\bibinfo {collaboration} {T2K}),\ }\bibfield  {title} {\bibinfo {title} {{The T2K Experiment}},\ }\href {https://doi.org/10.1016/j.nima.2011.06.067} {\bibfield  {journal} {\bibinfo  {journal} {Nucl. Instrum. Meth. A}\ }\textbf {\bibinfo {volume} {659}},\ \bibinfo {pages} {106} (\bibinfo {year} {2011})},\ \Eprint {https://arxiv.org/abs/1106.1238} {arXiv:1106.1238 [physics.ins-det]} \BibitemShut {NoStop}%
\bibitem [{\citenamefont {Cao}\ \emph {et~al.}(2014)\citenamefont {Cao} \emph {et~al.}}]{Cao:2014bea}%
  \BibitemOpen
  \bibfield  {author} {\bibinfo {author} {\bibfnamefont {J.}~\bibnamefont {Cao}} \emph {et~al.},\ }\bibfield  {title} {\bibinfo {title} {{Muon-decay medium-baseline neutrino beam facility}},\ }\href {https://doi.org/10.1103/PhysRevSTAB.17.090101} {\bibfield  {journal} {\bibinfo  {journal} {Phys. Rev. ST Accel. Beams}\ }\textbf {\bibinfo {volume} {17}},\ \bibinfo {pages} {090101} (\bibinfo {year} {2014})},\ \Eprint {https://arxiv.org/abs/1401.8125} {arXiv:1401.8125 [physics.acc-ph]} \BibitemShut {NoStop}%
\bibitem [{\citenamefont {Abe}\ \emph {et~al.}(2015)\citenamefont {Abe} \emph {et~al.}}]{Hyper-KamiokandeProto-:2015xww}%
  \BibitemOpen
  \bibfield  {author} {\bibinfo {author} {\bibfnamefont {K.}~\bibnamefont {Abe}} \emph {et~al.} (\bibinfo {collaboration} {Hyper-Kamiokande Proto-Collaboration}),\ }\bibfield  {title} {\bibinfo {title} {{Physics potential of a long-baseline neutrino oscillation experiment using a J-PARC neutrino beam and Hyper-Kamiokande}},\ }\href {https://doi.org/10.1093/ptep/ptv061} {\bibfield  {journal} {\bibinfo  {journal} {PTEP}\ }\textbf {\bibinfo {volume} {2015}},\ \bibinfo {pages} {053C02} (\bibinfo {year} {2015})},\ \Eprint {https://arxiv.org/abs/1502.05199} {arXiv:1502.05199 [hep-ex]} \BibitemShut {NoStop}%
\bibitem [{\citenamefont {Alekou}\ \emph {et~al.}(2021)\citenamefont {Alekou} \emph {et~al.}}]{ESSnuSB:2021azq}%
  \BibitemOpen
  \bibfield  {author} {\bibinfo {author} {\bibfnamefont {A.}~\bibnamefont {Alekou}} \emph {et~al.} (\bibinfo {collaboration} {ESSnuSB}),\ }\bibfield  {title} {\bibinfo {title} {{Updated physics performance of the ESSnuSB experiment: ESSnuSB collaboration}},\ }\href {https://doi.org/10.1140/epjc/s10052-021-09845-8} {\bibfield  {journal} {\bibinfo  {journal} {Eur. Phys. J. C}\ }\textbf {\bibinfo {volume} {81}},\ \bibinfo {pages} {1130} (\bibinfo {year} {2021})},\ \Eprint {https://arxiv.org/abs/2107.07585} {arXiv:2107.07585 [hep-ex]} \BibitemShut {NoStop}%
\bibitem [{\citenamefont {Nikolakopoulos}\ \emph {et~al.}(2021)\citenamefont {Nikolakopoulos}, \citenamefont {Pandey}, \citenamefont {Spitz},\ and\ \citenamefont {Jachowicz}}]{Nikolakopoulos:2020alk}%
  \BibitemOpen
  \bibfield  {author} {\bibinfo {author} {\bibfnamefont {A.}~\bibnamefont {Nikolakopoulos}}, \bibinfo {author} {\bibfnamefont {V.}~\bibnamefont {Pandey}}, \bibinfo {author} {\bibfnamefont {J.}~\bibnamefont {Spitz}},\ and\ \bibinfo {author} {\bibfnamefont {N.}~\bibnamefont {Jachowicz}},\ }\bibfield  {title} {\bibinfo {title} {{Modeling quasielastic interactions of monoenergetic kaon decay-at-rest neutrinos}},\ }\href {https://doi.org/10.1103/PhysRevC.103.064603} {\bibfield  {journal} {\bibinfo  {journal} {Phys. Rev. C}\ }\textbf {\bibinfo {volume} {103}},\ \bibinfo {pages} {064603} (\bibinfo {year} {2021})},\ \Eprint {https://arxiv.org/abs/2010.05794} {arXiv:2010.05794 [nucl-th]} \BibitemShut {NoStop}%
\bibitem [{\citenamefont {Rott}\ \emph {et~al.}(2015)\citenamefont {Rott}, \citenamefont {In}, \citenamefont {Kumar},\ and\ \citenamefont {Yaylali}}]{Rott:2015nma}%
  \BibitemOpen
  \bibfield  {author} {\bibinfo {author} {\bibfnamefont {C.}~\bibnamefont {Rott}}, \bibinfo {author} {\bibfnamefont {S.}~\bibnamefont {In}}, \bibinfo {author} {\bibfnamefont {J.}~\bibnamefont {Kumar}},\ and\ \bibinfo {author} {\bibfnamefont {D.}~\bibnamefont {Yaylali}},\ }\bibfield  {title} {\bibinfo {title} {{Dark Matter Searches for Monoenergetic Neutrinos Arising from Stopped Meson Decay in the Sun}},\ }\href {https://doi.org/10.1088/1475-7516/2015/11/039} {\bibfield  {journal} {\bibinfo  {journal} {Journ. of Cosm. and Astro. Phys.}\ }\textbf {\bibinfo {volume} {11}},\ \bibinfo {pages} {039} (\bibinfo {year} {2015})},\ \Eprint {https://arxiv.org/abs/1510.00170} {arXiv:1510.00170 [hep-ph]} \BibitemShut {NoStop}%
\bibitem [{\citenamefont {Abud}\ \emph {et~al.}(2021)\citenamefont {Abud} \emph {et~al.}}]{DUNE:2021gbm}%
  \BibitemOpen
  \bibfield  {author} {\bibinfo {author} {\bibfnamefont {A.~A.}\ \bibnamefont {Abud}} \emph {et~al.} (\bibinfo {collaboration} {DUNE}),\ }\bibfield  {title} {\bibinfo {title} {{Searching for solar KDAR with DUNE}},\ }\href {https://doi.org/10.1088/1475-7516/2021/10/065} {\bibfield  {journal} {\bibinfo  {journal} {Journ. of Cosm. and Astro. Phys.}\ }\textbf {\bibinfo {volume} {10}},\ \bibinfo {pages} {065} (\bibinfo {year} {2021})},\ \Eprint {https://arxiv.org/abs/2107.09109} {arXiv:2107.09109 [hep-ex]} \BibitemShut {NoStop}%
\bibitem [{\citenamefont {Spitz}(2012)}]{Spitz:2012gp}%
  \BibitemOpen
  \bibfield  {author} {\bibinfo {author} {\bibfnamefont {J.}~\bibnamefont {Spitz}},\ }\bibfield  {title} {\bibinfo {title} {{A Sterile Neutrino Search with Kaon Decay-at-rest}},\ }\href {https://doi.org/10.1103/PhysRevD.85.093020} {\bibfield  {journal} {\bibinfo  {journal} {Phys. Rev. D}\ }\textbf {\bibinfo {volume} {85}},\ \bibinfo {pages} {093020} (\bibinfo {year} {2012})},\ \Eprint {https://arxiv.org/abs/1203.6050} {arXiv:1203.6050 [hep-ph]} \BibitemShut {NoStop}%
\bibitem [{\citenamefont {Axani}\ \emph {et~al.}(2015)\citenamefont {Axani}, \citenamefont {Collin}, \citenamefont {Conrad}, \citenamefont {Shaevitz}, \citenamefont {Spitz},\ and\ \citenamefont {Wongjirad}}]{Axani:2015dha}%
  \BibitemOpen
  \bibfield  {author} {\bibinfo {author} {\bibfnamefont {S.}~\bibnamefont {Axani}}, \bibinfo {author} {\bibfnamefont {G.}~\bibnamefont {Collin}}, \bibinfo {author} {\bibfnamefont {J.~M.}\ \bibnamefont {Conrad}}, \bibinfo {author} {\bibfnamefont {M.~H.}\ \bibnamefont {Shaevitz}}, \bibinfo {author} {\bibfnamefont {J.}~\bibnamefont {Spitz}},\ and\ \bibinfo {author} {\bibfnamefont {T.}~\bibnamefont {Wongjirad}},\ }\bibfield  {title} {\bibinfo {title} {Decisive disappearance search at high $\mathrm{\ensuremath{\Delta}}{m}^{2}$ with monoenergetic muon neutrinos},\ }\href {https://doi.org/10.1103/PhysRevD.92.092010} {\bibfield  {journal} {\bibinfo  {journal} {Phys. Rev. D}\ }\textbf {\bibinfo {volume} {92}},\ \bibinfo {pages} {092010} (\bibinfo {year} {2015})}\BibitemShut {NoStop}%
\bibitem [{\citenamefont {Grant}\ and\ \citenamefont {Littlejohn}(2016)}]{Grant:2015jva}%
  \BibitemOpen
  \bibfield  {author} {\bibinfo {author} {\bibfnamefont {C.}~\bibnamefont {Grant}}\ and\ \bibinfo {author} {\bibfnamefont {B.}~\bibnamefont {Littlejohn}},\ }\bibfield  {title} {\bibinfo {title} {{Opportunities With Decay-At-Rest Neutrinos From Decay-In-Flight Neutrino Beams}},\ }\href {https://doi.org/10.22323/1.282.0483} {\bibfield  {journal} {\bibinfo  {journal} {PoS}\ }\textbf {\bibinfo {volume} {ICHEP2016}},\ \bibinfo {pages} {483} (\bibinfo {year} {2016})},\ \Eprint {https://arxiv.org/abs/1510.08431} {arXiv:1510.08431 [hep-ex]} \BibitemShut {NoStop}%
\bibitem [{\citenamefont {Harnik}\ \emph {et~al.}(2020)\citenamefont {Harnik}, \citenamefont {Kelly},\ and\ \citenamefont {Machado}}]{Harnik:2019iwv}%
  \BibitemOpen
  \bibfield  {author} {\bibinfo {author} {\bibfnamefont {R.}~\bibnamefont {Harnik}}, \bibinfo {author} {\bibfnamefont {K.~J.}\ \bibnamefont {Kelly}},\ and\ \bibinfo {author} {\bibfnamefont {P.~A.~N.}\ \bibnamefont {Machado}},\ }\bibfield  {title} {\bibinfo {title} {{Prospects of Measuring Oscillated Decay-at-Rest Neutrinos at Long Baselines}},\ }\href {https://doi.org/10.1103/PhysRevD.101.033008} {\bibfield  {journal} {\bibinfo  {journal} {Phys. Rev. D}\ }\textbf {\bibinfo {volume} {101}},\ \bibinfo {pages} {033008} (\bibinfo {year} {2020})},\ \Eprint {https://arxiv.org/abs/1911.05088} {arXiv:1911.05088 [hep-ph]} \BibitemShut {NoStop}%
\bibitem [{\citenamefont {Aguilar-Arevalo}\ \emph {et~al.}(2018)\citenamefont {Aguilar-Arevalo} \emph {et~al.}}]{MiniBooNE:2018dus}%
  \BibitemOpen
  \bibfield  {author} {\bibinfo {author} {\bibfnamefont {A.~A.}\ \bibnamefont {Aguilar-Arevalo}} \emph {et~al.} (\bibinfo {collaboration} {MiniBooNE}),\ }\bibfield  {title} {\bibinfo {title} {{First Measurement of Monoenergetic Muon Neutrino Charged Current Interactions}},\ }\href {https://doi.org/10.1103/PhysRevLett.120.141802} {\bibfield  {journal} {\bibinfo  {journal} {Phys. Rev. Lett.}\ }\textbf {\bibinfo {volume} {120}},\ \bibinfo {pages} {141802} (\bibinfo {year} {2018})},\ \Eprint {https://arxiv.org/abs/1801.03848} {arXiv:1801.03848 [hep-ex]} \BibitemShut {NoStop}%
\bibitem [{\citenamefont {Martini}\ \emph {et~al.}(2009)\citenamefont {Martini}, \citenamefont {Ericson}, \citenamefont {Chanfray},\ and\ \citenamefont {Marteau}}]{Martini:2009uj}%
  \BibitemOpen
  \bibfield  {author} {\bibinfo {author} {\bibfnamefont {M.}~\bibnamefont {Martini}}, \bibinfo {author} {\bibfnamefont {M.}~\bibnamefont {Ericson}}, \bibinfo {author} {\bibfnamefont {G.}~\bibnamefont {Chanfray}},\ and\ \bibinfo {author} {\bibfnamefont {J.}~\bibnamefont {Marteau}},\ }\bibfield  {title} {\bibinfo {title} {{A Unified approach for nucleon knock-out, coherent and incoherent pion production in neutrino interactions with nuclei}},\ }\href {https://doi.org/10.1103/PhysRevC.80.065501} {\bibfield  {journal} {\bibinfo  {journal} {Phys. Rev. C}\ }\textbf {\bibinfo {volume} {80}},\ \bibinfo {pages} {065501} (\bibinfo {year} {2009})},\ \Eprint {https://arxiv.org/abs/0910.2622} {arXiv:0910.2622 [nucl-th]} \BibitemShut {NoStop}%
\bibitem [{\citenamefont {Martini}\ \emph {et~al.}(2011)\citenamefont {Martini}, \citenamefont {Ericson},\ and\ \citenamefont {Chanfray}}]{Martini:2011wp}%
  \BibitemOpen
  \bibfield  {author} {\bibinfo {author} {\bibfnamefont {M.}~\bibnamefont {Martini}}, \bibinfo {author} {\bibfnamefont {M.}~\bibnamefont {Ericson}},\ and\ \bibinfo {author} {\bibfnamefont {G.}~\bibnamefont {Chanfray}},\ }\bibfield  {title} {\bibinfo {title} {{Neutrino quasielastic interaction and nuclear dynamics}},\ }\href {https://doi.org/10.1103/PhysRevC.84.055502} {\bibfield  {journal} {\bibinfo  {journal} {Phys. Rev. C}\ }\textbf {\bibinfo {volume} {84}},\ \bibinfo {pages} {055502} (\bibinfo {year} {2011})},\ \Eprint {https://arxiv.org/abs/1110.0221} {arXiv:1110.0221 [nucl-th]} \BibitemShut {NoStop}%
\bibitem [{\citenamefont {Akbar}\ \emph {et~al.}(2017)\citenamefont {Akbar}, \citenamefont {Sajjad~Athar},\ and\ \citenamefont {Singh}}]{KDAR-RPA-2017}%
  \BibitemOpen
  \bibfield  {author} {\bibinfo {author} {\bibfnamefont {F.}~\bibnamefont {Akbar}}, \bibinfo {author} {\bibfnamefont {M.}~\bibnamefont {Sajjad~Athar}},\ and\ \bibinfo {author} {\bibfnamefont {S.~K.}\ \bibnamefont {Singh}},\ }\bibfield  {title} {\bibinfo {title} {{Neutrino-nucleus cross sections in $^{12}C$ and $^{40}Ar$ with KDAR neutrinos}},\ }\href {https://doi.org/10.1088/1361-6471/aa9125} {\bibfield  {journal} {\bibinfo  {journal} {J. Phys. G}\ }\textbf {\bibinfo {volume} {44}},\ \bibinfo {pages} {125108} (\bibinfo {year} {2017})},\ \Eprint {https://arxiv.org/abs/1708.00321} {arXiv:1708.00321 [nucl-th]} \BibitemShut {NoStop}%
\bibitem [{\citenamefont {Agostinelli}\ \emph {et~al.}(2003)\citenamefont {Agostinelli} \emph {et~al.}}]{Geant4}%
  \BibitemOpen
  \bibfield  {author} {\bibinfo {author} {\bibfnamefont {S.}~\bibnamefont {Agostinelli}} \emph {et~al.} (\bibinfo {collaboration} {GEANT4}),\ }\bibfield  {title} {\bibinfo {title} {{GEANT4--a simulation toolkit}},\ }\href {https://doi.org/10.1016/S0168-9002(03)01368-8} {\bibfield  {journal} {\bibinfo  {journal} {Nucl. Instrum. Meth. A}\ }\textbf {\bibinfo {volume} {506}},\ \bibinfo {pages} {250} (\bibinfo {year} {2003})}\BibitemShut {NoStop}%
\bibitem [{\citenamefont {{N.V. Mokhov, FERMILAB-FN-628, (1995); O.E. Krivosheev and N.V. Mokhov, “MARS Code Status”, Fermilab-Conf-00/181, (2000); O.E. Krivosheev and N.V. Mokhov, “Status of MARS Code”, Fermilab-Conf03/053, (2003); N.V. Mokhov, K.K. Gudima, C.C. James \textit{et al.}, “Recent Enhancements to the MARS15 Code”, Fermilab-Conf-04/053}}(2004)}]{osti_1282121}%
  \BibitemOpen
  \bibfield  {author} {\bibinfo {author} {\bibnamefont {{N.V. Mokhov, FERMILAB-FN-628, (1995); O.E. Krivosheev and N.V. Mokhov, “MARS Code Status”, Fermilab-Conf-00/181, (2000); O.E. Krivosheev and N.V. Mokhov, “Status of MARS Code”, Fermilab-Conf03/053, (2003); N.V. Mokhov, K.K. Gudima, C.C. James \textit{et al.}, “Recent Enhancements to the MARS15 Code”, Fermilab-Conf-04/053}}},\ }\href@noop {} {} (\bibinfo {year} {2004})\BibitemShut {NoStop}%
\bibitem [{\citenamefont {Harada}\ \emph {et~al.}(2013)\citenamefont {Harada} \emph {et~al.}}]{JSNS2:2013jdh}%
  \BibitemOpen
  \bibfield  {author} {\bibinfo {author} {\bibfnamefont {M.}~\bibnamefont {Harada}} \emph {et~al.} (\bibinfo {collaboration} {JSNS2}),\ }\bibfield  {title} {\bibinfo {title} {{Proposal: A Search for Sterile Neutrino at J-PARC Materials and Life Science Experimental Facility}},\ }\href@noop {} {\  (\bibinfo {year} {2013})},\ \Eprint {https://arxiv.org/abs/1310.1437} {arXiv:1310.1437 [physics.ins-det]} \BibitemShut {NoStop}%
\bibitem [{\citenamefont {Ajimura}\ \emph {et~al.}(2017)\citenamefont {Ajimura} \emph {et~al.}}]{Ajimura:2017fld}%
  \BibitemOpen
  \bibfield  {author} {\bibinfo {author} {\bibfnamefont {S.}~\bibnamefont {Ajimura}} \emph {et~al.},\ }\bibfield  {title} {\bibinfo {title} {{Technical Design Report (TDR): Searching for a Sterile Neutrino at J-PARC MLF (E56, JSNS2)}},\ }\href@noop {} {\  (\bibinfo {year} {2017})},\ \Eprint {https://arxiv.org/abs/1705.08629} {arXiv:1705.08629 [physics.ins-det]} \BibitemShut {NoStop}%
\bibitem [{\citenamefont {Ajimura}\ \emph {et~al.}(2021)\citenamefont {Ajimura} \emph {et~al.}}]{JSNS2:2021hyk}%
  \BibitemOpen
  \bibfield  {author} {\bibinfo {author} {\bibfnamefont {S.}~\bibnamefont {Ajimura}} \emph {et~al.} (\bibinfo {collaboration} {JSNS2}),\ }\bibfield  {title} {\bibinfo {title} {{The JSNS2 detector}},\ }\href {https://doi.org/10.1016/j.nima.2021.165742} {\bibfield  {journal} {\bibinfo  {journal} {Nucl. Instrum. Meth. A}\ }\textbf {\bibinfo {volume} {1014}},\ \bibinfo {pages} {165742} (\bibinfo {year} {2021})},\ \Eprint {https://arxiv.org/abs/2104.13169} {arXiv:2104.13169 [physics.ins-det]} \BibitemShut {NoStop}%
\bibitem [{\citenamefont {Makins}\ \emph {et~al.}(1994)\citenamefont {Makins} \emph {et~al.}}]{Makins:1994mm}%
  \BibitemOpen
  \bibfield  {author} {\bibinfo {author} {\bibfnamefont {N.}~\bibnamefont {Makins}} \emph {et~al.},\ }\bibfield  {title} {\bibinfo {title} {{Momentum transfer dependence of nuclear transparency from the quasielastic C-12 (e, e-prime p) reaction}},\ }\href {https://doi.org/10.1103/PhysRevLett.72.1986} {\bibfield  {journal} {\bibinfo  {journal} {Phys. Rev. Lett.}\ }\textbf {\bibinfo {volume} {72}},\ \bibinfo {pages} {1986} (\bibinfo {year} {1994})}\BibitemShut {NoStop}%
\bibitem [{\citenamefont {Eckhause}\ \emph {et~al.}(1963)\citenamefont {Eckhause}, \citenamefont {Filippas}, \citenamefont {Sutton},\ and\ \citenamefont {Welsh}}]{MuonLifetimePaper}%
  \BibitemOpen
  \bibfield  {author} {\bibinfo {author} {\bibfnamefont {M.}~\bibnamefont {Eckhause}}, \bibinfo {author} {\bibfnamefont {T.~A.}\ \bibnamefont {Filippas}}, \bibinfo {author} {\bibfnamefont {R.~B.}\ \bibnamefont {Sutton}},\ and\ \bibinfo {author} {\bibfnamefont {R.~E.}\ \bibnamefont {Welsh}},\ }\bibfield  {title} {\bibinfo {title} {{Measurement of Negative muon lifetime in Light Isotopes}},\ }\href {https://doi.org/10.1103/PhysRev.132.422} {\bibfield  {journal} {\bibinfo  {journal} {Phys. Rev.}\ }\textbf {\bibinfo {volume} {132}},\ \bibinfo {pages} {422} (\bibinfo {year} {1963})}\BibitemShut {NoStop}%
\bibitem [{\citenamefont {Jordan}(2022)}]{jrjordan_thesis}%
  \BibitemOpen
  \bibfield  {author} {\bibinfo {author} {\bibfnamefont {J.}~\bibnamefont {Jordan}},\ }\emph {\bibinfo {title} {Physics at Short Baseline Neutrino Experiments}},\ \href {https://doi.org/https://dx.doi.org/10.7302/6288} {Ph.D. thesis},\ \bibinfo  {school} {University of Michigan} (\bibinfo {year} {2022})\BibitemShut {NoStop}%
\bibitem [{\citenamefont {Lee}\ \emph {et~al.}(2024)\citenamefont {Lee} \emph {et~al.}}]{Lee:2024pwg}%
  \BibitemOpen
  \bibfield  {author} {\bibinfo {author} {\bibfnamefont {D.~H.}\ \bibnamefont {Lee}} \emph {et~al.},\ }\bibfield  {title} {\bibinfo {title} {{Evaluation of the performance of the event reconstruction algorithms in the JSNS$^2$ experiment using a $^{252}$Cf calibration source}},\ }\href@noop {} {\  (\bibinfo {year} {2024})},\ \Eprint {https://arxiv.org/abs/2404.04153} {arXiv:2404.04153 [hep-ex]} \BibitemShut {NoStop}%
\bibitem [{\citenamefont {An}\ \emph {et~al.}(2012)\citenamefont {An} \emph {et~al.}}]{DayaBay:2012fng}%
  \BibitemOpen
  \bibfield  {author} {\bibinfo {author} {\bibfnamefont {F.~P.}\ \bibnamefont {An}} \emph {et~al.} (\bibinfo {collaboration} {Daya Bay}),\ }\bibfield  {title} {\bibinfo {title} {{Observation of electron-antineutrino disappearance at Daya Bay}},\ }\href {https://doi.org/10.1103/PhysRevLett.108.171803} {\bibfield  {journal} {\bibinfo  {journal} {Phys. Rev. Lett.}\ }\textbf {\bibinfo {volume} {108}},\ \bibinfo {pages} {171803} (\bibinfo {year} {2012})},\ \Eprint {https://arxiv.org/abs/1203.1669} {arXiv:1203.1669 [hep-ex]} \BibitemShut {NoStop}%
\bibitem [{\citenamefont {Ahn}\ \emph {et~al.}(2012)\citenamefont {Ahn} \emph {et~al.}}]{RENO:2012mkc}%
  \BibitemOpen
  \bibfield  {author} {\bibinfo {author} {\bibfnamefont {J.~K.}\ \bibnamefont {Ahn}} \emph {et~al.} (\bibinfo {collaboration} {RENO}),\ }\bibfield  {title} {\bibinfo {title} {{Observation of Reactor Electron Antineutrino Disappearance in the RENO Experiment}},\ }\href {https://doi.org/10.1103/PhysRevLett.108.191802} {\bibfield  {journal} {\bibinfo  {journal} {Phys. Rev. Lett.}\ }\textbf {\bibinfo {volume} {108}},\ \bibinfo {pages} {191802} (\bibinfo {year} {2012})},\ \Eprint {https://arxiv.org/abs/1204.0626} {arXiv:1204.0626 [hep-ex]} \BibitemShut {NoStop}%
\bibitem [{\citenamefont {Golan}\ \emph {et~al.}(2012{\natexlab{a}})\citenamefont {Golan}, \citenamefont {Juszczak},\ and\ \citenamefont {Sobczyk}}]{Golan:2012wx}%
  \BibitemOpen
  \bibfield  {author} {\bibinfo {author} {\bibfnamefont {T.}~\bibnamefont {Golan}}, \bibinfo {author} {\bibfnamefont {C.}~\bibnamefont {Juszczak}},\ and\ \bibinfo {author} {\bibfnamefont {J.~T.}\ \bibnamefont {Sobczyk}},\ }\bibfield  {title} {\bibinfo {title} {{Final State Interactions Effects in Neutrino-Nucleus Interactions}},\ }\href {https://doi.org/10.1103/PhysRevC.86.015505} {\bibfield  {journal} {\bibinfo  {journal} {Phys. Rev. C}\ }\textbf {\bibinfo {volume} {86}},\ \bibinfo {pages} {015505} (\bibinfo {year} {2012}{\natexlab{a}})},\ \Eprint {https://arxiv.org/abs/1202.4197} {arXiv:1202.4197 [nucl-th]} \BibitemShut {NoStop}%
\bibitem [{\citenamefont {Golan}\ \emph {et~al.}(2012{\natexlab{b}})\citenamefont {Golan}, \citenamefont {Sobczyk},\ and\ \citenamefont {Zmuda}}]{Golan:2012rfa}%
  \BibitemOpen
  \bibfield  {author} {\bibinfo {author} {\bibfnamefont {T.}~\bibnamefont {Golan}}, \bibinfo {author} {\bibfnamefont {J.~T.}\ \bibnamefont {Sobczyk}},\ and\ \bibinfo {author} {\bibfnamefont {J.}~\bibnamefont {Zmuda}},\ }\bibfield  {title} {\bibinfo {title} {{NuWro: the Wroclaw Monte Carlo Generator of Neutrino Interactions}},\ }\href {https://doi.org/10.1016/j.nuclphysbps.2012.09.136} {\bibfield  {journal} {\bibinfo  {journal} {Nucl. Phys. B Proc. Suppl.}\ }\textbf {\bibinfo {volume} {229-232}},\ \bibinfo {pages} {499} (\bibinfo {year} {2012}{\natexlab{b}})}\BibitemShut {NoStop}%
\bibitem [{\citenamefont {Lalakulich}\ \emph {et~al.}(2013)\citenamefont {Lalakulich}, \citenamefont {Gallmeister},\ and\ \citenamefont {Mosel}}]{Lalakulich:2011eh}%
  \BibitemOpen
  \bibfield  {author} {\bibinfo {author} {\bibfnamefont {O.}~\bibnamefont {Lalakulich}}, \bibinfo {author} {\bibfnamefont {K.}~\bibnamefont {Gallmeister}},\ and\ \bibinfo {author} {\bibfnamefont {U.}~\bibnamefont {Mosel}},\ }\bibfield  {title} {\bibinfo {title} {{Neutrino Nucleus Reactions within the GiBUU Model}},\ }\href {https://doi.org/10.1088/1742-6596/408/1/012053} {\bibfield  {journal} {\bibinfo  {journal} {J. Phys. Conf. Ser.}\ }\textbf {\bibinfo {volume} {408}},\ \bibinfo {pages} {012053} (\bibinfo {year} {2013})},\ \Eprint {https://arxiv.org/abs/1110.0674} {arXiv:1110.0674 [hep-ph]} \BibitemShut {NoStop}%
\bibitem [{\citenamefont {Buss}\ \emph {et~al.}(2012)\citenamefont {Buss}, \citenamefont {Gaitanos}, \citenamefont {Gallmeister}, \citenamefont {van Hees}, \citenamefont {Kaskulov}, \citenamefont {Lalakulich}, \citenamefont {Larionov}, \citenamefont {Leitner}, \citenamefont {Weil},\ and\ \citenamefont {Mosel}}]{Buss:2011mx}%
  \BibitemOpen
  \bibfield  {author} {\bibinfo {author} {\bibfnamefont {O.}~\bibnamefont {Buss}}, \bibinfo {author} {\bibfnamefont {T.}~\bibnamefont {Gaitanos}}, \bibinfo {author} {\bibfnamefont {K.}~\bibnamefont {Gallmeister}}, \bibinfo {author} {\bibfnamefont {H.}~\bibnamefont {van Hees}}, \bibinfo {author} {\bibfnamefont {M.}~\bibnamefont {Kaskulov}}, \bibinfo {author} {\bibfnamefont {O.}~\bibnamefont {Lalakulich}}, \bibinfo {author} {\bibfnamefont {A.~B.}\ \bibnamefont {Larionov}}, \bibinfo {author} {\bibfnamefont {T.}~\bibnamefont {Leitner}}, \bibinfo {author} {\bibfnamefont {J.}~\bibnamefont {Weil}},\ and\ \bibinfo {author} {\bibfnamefont {U.}~\bibnamefont {Mosel}},\ }\bibfield  {title} {\bibinfo {title} {{Transport-theoretical Description of Nuclear Reactions}},\ }\href {https://doi.org/10.1016/j.physrep.2011.12.001} {\bibfield  {journal} {\bibinfo  {journal} {Phys. Rept.}\ }\textbf {\bibinfo {volume} {512}},\ \bibinfo {pages} {1} (\bibinfo {year} {2012})},\ \Eprint {https://arxiv.org/abs/1106.1344} {arXiv:1106.1344
  [hep-ph]} \BibitemShut {NoStop}%
\bibitem [{\citenamefont {Abe}(2024)}]{NucDeEx}%
  \BibitemOpen
  \bibfield  {author} {\bibinfo {author} {\bibfnamefont {S.}~\bibnamefont {Abe}},\ }\bibfield  {title} {\bibinfo {title} {Nuclear deexcitation simulator for neutrino interactions and nucleon decays of $^{12}\mathrm{C}$ and $^{16}\mathrm{O}$ based on talys},\ }\href {https://doi.org/10.1103/PhysRevD.109.036009} {\bibfield  {journal} {\bibinfo  {journal} {Phys. Rev. D}\ }\textbf {\bibinfo {volume} {109}},\ \bibinfo {pages} {036009} (\bibinfo {year} {2024})}\BibitemShut {NoStop}%
\bibitem [{\citenamefont {Koning}\ \emph {et~al.}(2023)\citenamefont {Koning}, \citenamefont {Hilaire},\ and\ \citenamefont {Goriely}}]{TALYS}%
  \BibitemOpen
  \bibfield  {author} {\bibinfo {author} {\bibfnamefont {A.}~\bibnamefont {Koning}}, \bibinfo {author} {\bibfnamefont {S.}~\bibnamefont {Hilaire}},\ and\ \bibinfo {author} {\bibfnamefont {S.}~\bibnamefont {Goriely}},\ }\bibfield  {title} {\bibinfo {title} {Talys: modeling of nuclear reactions},\ }\href {https://doi.org/10.1140/epja/s10050-023-01034-3} {\bibfield  {journal} {\bibinfo  {journal} {The European Physical Journal A}\ }\textbf {\bibinfo {volume} {59}},\ \bibinfo {pages} {131} (\bibinfo {year} {2023})}\BibitemShut {NoStop}%
\bibitem [{\citenamefont {Jeon}(2022)}]{hyoungku_thesis}%
  \BibitemOpen
  \bibfield  {author} {\bibinfo {author} {\bibfnamefont {H.~K.}\ \bibnamefont {Jeon}},\ }\emph {\bibinfo {title} {Search for neutrinos from charged kaon decay-at-rest using JSNS2 first year of data with J-PARC MLF 3 GeV proton beam}},\ \href {https://dcollection.skku.edu/srch/srchDetail/000000171207} {Ph.D. thesis},\ \bibinfo  {school} {Sungkyunkwan University} (\bibinfo {year} {2022})\BibitemShut {NoStop}%
\bibitem [{\citenamefont {Wright}\ and\ \citenamefont {Kelsey}(2015)}]{Wright:2015xia}%
  \BibitemOpen
  \bibfield  {author} {\bibinfo {author} {\bibfnamefont {D.~H.}\ \bibnamefont {Wright}}\ and\ \bibinfo {author} {\bibfnamefont {M.~H.}\ \bibnamefont {Kelsey}},\ }\bibfield  {title} {\bibinfo {title} {{The Geant4 Bertini Cascade}},\ }\href {https://doi.org/10.1016/j.nima.2015.09.058} {\bibfield  {journal} {\bibinfo  {journal} {Nucl. Instrum. Meth. A}\ }\textbf {\bibinfo {volume} {804}},\ \bibinfo {pages} {175} (\bibinfo {year} {2015})}\BibitemShut {NoStop}%
\bibitem [{\citenamefont {Folger}\ \emph {et~al.}(2004)\citenamefont {Folger}, \citenamefont {Ivanchenko},\ and\ \citenamefont {Wellisch}}]{binary_cascade}%
  \BibitemOpen
  \bibfield  {author} {\bibinfo {author} {\bibfnamefont {G.}~\bibnamefont {Folger}}, \bibinfo {author} {\bibfnamefont {V.~N.}\ \bibnamefont {Ivanchenko}},\ and\ \bibinfo {author} {\bibfnamefont {J.~P.}\ \bibnamefont {Wellisch}},\ }\bibfield  {title} {\bibinfo {title} {The binary cascade},\ }\href {https://doi.org/10.1140/epja/i2003-10219-7} {\bibfield  {journal} {\bibinfo  {journal} {The European Physical Journal A - Hadrons and Nuclei}\ }\textbf {\bibinfo {volume} {21}},\ \bibinfo {pages} {407} (\bibinfo {year} {2004})}\BibitemShut {NoStop}%
\bibitem [{\citenamefont {Shepp}\ and\ \citenamefont {Vardi}(1982)}]{iterative_unfolding1}%
  \BibitemOpen
  \bibfield  {author} {\bibinfo {author} {\bibfnamefont {L.~A.}\ \bibnamefont {Shepp}}\ and\ \bibinfo {author} {\bibfnamefont {Y.}~\bibnamefont {Vardi}},\ }\bibfield  {title} {\bibinfo {title} {Maximum likelihood reconstruction for emission tomography},\ }\href {https://doi.org/10.1109/TMI.1982.4307558} {\bibfield  {journal} {\bibinfo  {journal} {IEEE Transactions on Medical Imaging}\ }\textbf {\bibinfo {volume} {1}},\ \bibinfo {pages} {113} (\bibinfo {year} {1982})}\BibitemShut {NoStop}%
\bibitem [{\citenamefont {D'Agostini}(1995)}]{dagostini_unfolding}%
  \BibitemOpen
  \bibfield  {author} {\bibinfo {author} {\bibfnamefont {G.}~\bibnamefont {D'Agostini}},\ }\bibfield  {title} {\bibinfo {title} {A multidimensional unfolding method based on bayes' theorem},\ }\href {https://doi.org/https://doi.org/10.1016/0168-9002(95)00274-X} {\bibfield  {journal} {\bibinfo  {journal} {Nuclear Instruments and Methods in Physics Research Section A: Accelerators, Spectrometers, Detectors and Associated Equipment}\ }\textbf {\bibinfo {volume} {362}},\ \bibinfo {pages} {487} (\bibinfo {year} {1995})}\BibitemShut {NoStop}%
\bibitem [{\citenamefont {Bourbeau}\ and\ \citenamefont {Hampel-Arias}(2018)}]{pyunfold}%
  \BibitemOpen
  \bibfield  {author} {\bibinfo {author} {\bibfnamefont {J.}~\bibnamefont {Bourbeau}}\ and\ \bibinfo {author} {\bibfnamefont {Z.}~\bibnamefont {Hampel-Arias}},\ }\bibfield  {title} {\bibinfo {title} {Pyunfold: A python package for iterative unfolding},\ }\href {https://doi.org/10.21105/joss.00741} {\bibfield  {journal} {\bibinfo  {journal} {The Journal of Open Source Software}\ }\textbf {\bibinfo {volume} {3}},\ \bibinfo {pages} {741} (\bibinfo {year} {2018})}\BibitemShut {NoStop}%
\bibitem [{\citenamefont {von Krosigk}\ \emph {et~al.}(2013)\citenamefont {von Krosigk}, \citenamefont {Neumann}, \citenamefont {Nolte}, \citenamefont {R\"ottger},\ and\ \citenamefont {Zuber}}]{vonKrosigk:2013sa}%
  \BibitemOpen
  \bibfield  {author} {\bibinfo {author} {\bibfnamefont {B.}~\bibnamefont {von Krosigk}}, \bibinfo {author} {\bibfnamefont {L.}~\bibnamefont {Neumann}}, \bibinfo {author} {\bibfnamefont {R.}~\bibnamefont {Nolte}}, \bibinfo {author} {\bibfnamefont {S.}~\bibnamefont {R\"ottger}},\ and\ \bibinfo {author} {\bibfnamefont {K.}~\bibnamefont {Zuber}},\ }\bibfield  {title} {\bibinfo {title} {{Measurement of the proton light response of various LAB based scintillators and its implication for supernova neutrino detection via neutrino-proton scattering}},\ }\href {https://doi.org/10.1140/epjc/s10052-013-2390-1} {\bibfield  {journal} {\bibinfo  {journal} {Eur. Phys. J. C}\ }\textbf {\bibinfo {volume} {73}},\ \bibinfo {pages} {2390} (\bibinfo {year} {2013})},\ \Eprint {https://arxiv.org/abs/1301.6403} {arXiv:1301.6403 [astro-ph.IM]} \BibitemShut {NoStop}%
\bibitem [{\citenamefont {Laplace}\ \emph {et~al.}(2022)\citenamefont {Laplace}, \citenamefont {Goldblum}, \citenamefont {Brown}, \citenamefont {LeBlanc}, \citenamefont {Li}, \citenamefont {Manfredi},\ and\ \citenamefont {Brubaker}}]{din_birks}%
  \BibitemOpen
  \bibfield  {author} {\bibinfo {author} {\bibfnamefont {T.~A.}\ \bibnamefont {Laplace}}, \bibinfo {author} {\bibfnamefont {B.~L.}\ \bibnamefont {Goldblum}}, \bibinfo {author} {\bibfnamefont {J.~A.}\ \bibnamefont {Brown}}, \bibinfo {author} {\bibfnamefont {G.}~\bibnamefont {LeBlanc}}, \bibinfo {author} {\bibfnamefont {T.}~\bibnamefont {Li}}, \bibinfo {author} {\bibfnamefont {J.~J.}\ \bibnamefont {Manfredi}},\ and\ \bibinfo {author} {\bibfnamefont {E.}~\bibnamefont {Brubaker}},\ }\bibfield  {title} {\bibinfo {title} {Modeling ionization quenching in organic scintillators},\ }\href {https://doi.org/10.1039/D2MA00388K} {\bibfield  {journal} {\bibinfo  {journal} {Mater. Adv.}\ }\textbf {\bibinfo {volume} {3}},\ \bibinfo {pages} {5871} (\bibinfo {year} {2022})}\BibitemShut {NoStop}%
\bibitem [{\citenamefont {Gonz\'alez-Jim\'enez}\ \emph {et~al.}(2019)\citenamefont {Gonz\'alez-Jim\'enez}, \citenamefont {Nikolakopoulos}, \citenamefont {Jachowicz},\ and\ \citenamefont {Ud\'{\i}as}}]{rmf_simulation}%
  \BibitemOpen
  \bibfield  {author} {\bibinfo {author} {\bibfnamefont {R.}~\bibnamefont {Gonz\'alez-Jim\'enez}}, \bibinfo {author} {\bibfnamefont {A.}~\bibnamefont {Nikolakopoulos}}, \bibinfo {author} {\bibfnamefont {N.}~\bibnamefont {Jachowicz}},\ and\ \bibinfo {author} {\bibfnamefont {J.~M.}\ \bibnamefont {Ud\'{\i}as}},\ }\bibfield  {title} {\bibinfo {title} {Nuclear effects in electron-nucleus and neutrino-nucleus scattering within a relativistic quantum mechanical framework},\ }\href {https://doi.org/10.1103/PhysRevC.100.045501} {\bibfield  {journal} {\bibinfo  {journal} {Phys. Rev. C}\ }\textbf {\bibinfo {volume} {100}},\ \bibinfo {pages} {045501} (\bibinfo {year} {2019})}\BibitemShut {NoStop}%
\bibitem [{\citenamefont {Benhar}\ \emph {et~al.}(1994)\citenamefont {Benhar}, \citenamefont {Fabrocini}, \citenamefont {Fantoni},\ and\ \citenamefont {Sick}}]{rmf_spectral_function}%
  \BibitemOpen
  \bibfield  {author} {\bibinfo {author} {\bibfnamefont {O.}~\bibnamefont {Benhar}}, \bibinfo {author} {\bibfnamefont {A.}~\bibnamefont {Fabrocini}}, \bibinfo {author} {\bibfnamefont {S.}~\bibnamefont {Fantoni}},\ and\ \bibinfo {author} {\bibfnamefont {I.}~\bibnamefont {Sick}},\ }\bibfield  {title} {\bibinfo {title} {Spectral function of finite nuclei and scattering of gev electrons},\ }\href {https://doi.org/https://doi.org/10.1016/0375-9474(94)90920-2} {\bibfield  {journal} {\bibinfo  {journal} {Nuclear Physics A}\ }\textbf {\bibinfo {volume} {579}},\ \bibinfo {pages} {493} (\bibinfo {year} {1994})}\BibitemShut {NoStop}%
\bibitem [{\citenamefont {Gonz\'alez-Jim\'enez}\ \emph {et~al.}(2022)\citenamefont {Gonz\'alez-Jim\'enez}, \citenamefont {Barbaro}, \citenamefont {Caballero}, \citenamefont {Donnelly}, \citenamefont {Jachowicz}, \citenamefont {Megias}, \citenamefont {Niewczas}, \citenamefont {Nikolakopoulos}, \citenamefont {Van~Orden},\ and\ \citenamefont {Ud\'{\i}as}}]{PhysRevC.105.025502}%
  \BibitemOpen
  \bibfield  {author} {\bibinfo {author} {\bibfnamefont {R.}~\bibnamefont {Gonz\'alez-Jim\'enez}}, \bibinfo {author} {\bibfnamefont {M.~B.}\ \bibnamefont {Barbaro}}, \bibinfo {author} {\bibfnamefont {J.~A.}\ \bibnamefont {Caballero}}, \bibinfo {author} {\bibfnamefont {T.~W.}\ \bibnamefont {Donnelly}}, \bibinfo {author} {\bibfnamefont {N.}~\bibnamefont {Jachowicz}}, \bibinfo {author} {\bibfnamefont {G.~D.}\ \bibnamefont {Megias}}, \bibinfo {author} {\bibfnamefont {K.}~\bibnamefont {Niewczas}}, \bibinfo {author} {\bibfnamefont {A.}~\bibnamefont {Nikolakopoulos}}, \bibinfo {author} {\bibfnamefont {J.~W.}\ \bibnamefont {Van~Orden}},\ and\ \bibinfo {author} {\bibfnamefont {J.~M.}\ \bibnamefont {Ud\'{\i}as}},\ }\bibfield  {title} {\bibinfo {title} {Neutrino energy reconstruction from semi-inclusive samples},\ }\href {https://doi.org/10.1103/PhysRevC.105.025502} {\bibfield  {journal} {\bibinfo  {journal} {Phys. Rev. C}\ }\textbf {\bibinfo {volume} {105}},\ \bibinfo {pages} {025502} (\bibinfo {year}
  {2022})}\BibitemShut {NoStop}%
\bibitem [{\citenamefont {Isaacson}\ \emph {et~al.}(2023)\citenamefont {Isaacson}, \citenamefont {Jay}, \citenamefont {Lovato}, \citenamefont {Machado},\ and\ \citenamefont {Rocco}}]{achilles}%
  \BibitemOpen
  \bibfield  {author} {\bibinfo {author} {\bibfnamefont {J.}~\bibnamefont {Isaacson}}, \bibinfo {author} {\bibfnamefont {W.~I.}\ \bibnamefont {Jay}}, \bibinfo {author} {\bibfnamefont {A.}~\bibnamefont {Lovato}}, \bibinfo {author} {\bibfnamefont {P.~A.~N.}\ \bibnamefont {Machado}},\ and\ \bibinfo {author} {\bibfnamefont {N.}~\bibnamefont {Rocco}},\ }\bibfield  {title} {\bibinfo {title} {Introducing a novel event generator for electron-nucleus and neutrino-nucleus scattering},\ }\href {https://doi.org/10.1103/PhysRevD.107.033007} {\bibfield  {journal} {\bibinfo  {journal} {Phys. Rev. D}\ }\textbf {\bibinfo {volume} {107}},\ \bibinfo {pages} {033007} (\bibinfo {year} {2023})}\BibitemShut {NoStop}%
\bibitem [{\citenamefont {Isaacson}\ \emph {et~al.}(2021)\citenamefont {Isaacson}, \citenamefont {Jay}, \citenamefont {Lovato}, \citenamefont {Machado},\ and\ \citenamefont {Rocco}}]{achilles_inc}%
  \BibitemOpen
  \bibfield  {author} {\bibinfo {author} {\bibfnamefont {J.}~\bibnamefont {Isaacson}}, \bibinfo {author} {\bibfnamefont {W.~I.}\ \bibnamefont {Jay}}, \bibinfo {author} {\bibfnamefont {A.}~\bibnamefont {Lovato}}, \bibinfo {author} {\bibfnamefont {P.~A.~N.}\ \bibnamefont {Machado}},\ and\ \bibinfo {author} {\bibfnamefont {N.}~\bibnamefont {Rocco}},\ }\bibfield  {title} {\bibinfo {title} {New approach to intranuclear cascades with quantum monte carlo configurations},\ }\href {https://doi.org/10.1103/PhysRevC.103.015502} {\bibfield  {journal} {\bibinfo  {journal} {Phys. Rev. C}\ }\textbf {\bibinfo {volume} {103}},\ \bibinfo {pages} {015502} (\bibinfo {year} {2021})}\BibitemShut {NoStop}%
\bibitem [{ale()}]{alexis_nikolakopoulos_pc}%
  \BibitemOpen
  \href@noop {} {}\bibinfo {note} {Alexis Nikolakopoulos (anikolak@fnal.gov), private communications}\BibitemShut {NoStop}%
\bibitem [{\citenamefont {Chakrani}\ \emph {et~al.}(2024)\citenamefont {Chakrani}, \citenamefont {Dolan}, \citenamefont {Avanzini}, \citenamefont {Ershova}, \citenamefont {Koch}, \citenamefont {McFarland}, \citenamefont {Megias}, \citenamefont {Munteanu}, \citenamefont {Pickering}, \citenamefont {Skwarczynski}, \citenamefont {Nguyen},\ and\ \citenamefont {Wret}}]{PhysRevD.109.072006}%
  \BibitemOpen
  \bibfield  {author} {\bibinfo {author} {\bibfnamefont {J.}~\bibnamefont {Chakrani}}, \bibinfo {author} {\bibfnamefont {S.}~\bibnamefont {Dolan}}, \bibinfo {author} {\bibfnamefont {M.~B.}\ \bibnamefont {Avanzini}}, \bibinfo {author} {\bibfnamefont {A.}~\bibnamefont {Ershova}}, \bibinfo {author} {\bibfnamefont {L.}~\bibnamefont {Koch}}, \bibinfo {author} {\bibfnamefont {K.}~\bibnamefont {McFarland}}, \bibinfo {author} {\bibfnamefont {G.~D.}\ \bibnamefont {Megias}}, \bibinfo {author} {\bibfnamefont {L.}~\bibnamefont {Munteanu}}, \bibinfo {author} {\bibfnamefont {L.}~\bibnamefont {Pickering}}, \bibinfo {author} {\bibfnamefont {K.}~\bibnamefont {Skwarczynski}}, \bibinfo {author} {\bibfnamefont {V.~Q.}\ \bibnamefont {Nguyen}},\ and\ \bibinfo {author} {\bibfnamefont {C.}~\bibnamefont {Wret}},\ }\bibfield  {title} {\bibinfo {title} {Parametrized uncertainties in the spectral function model of neutrino charged-current quasielastic interactions for oscillation analyses},\ }\href
  {https://doi.org/10.1103/PhysRevD.109.072006} {\bibfield  {journal} {\bibinfo  {journal} {Phys. Rev. D}\ }\textbf {\bibinfo {volume} {109}},\ \bibinfo {pages} {072006} (\bibinfo {year} {2024})}\BibitemShut {NoStop}%
\bibitem [{\citenamefont {Ankowski}\ \emph {et~al.}(2024)\citenamefont {Ankowski}, \citenamefont {Benhar},\ and\ \citenamefont {Sakuda}}]{ankowski_sf}%
  \BibitemOpen
  \bibfield  {author} {\bibinfo {author} {\bibfnamefont {A.~M.}\ \bibnamefont {Ankowski}}, \bibinfo {author} {\bibfnamefont {O.}~\bibnamefont {Benhar}},\ and\ \bibinfo {author} {\bibfnamefont {M.}~\bibnamefont {Sakuda}},\ }\href {https://arxiv.org/abs/2407.18226} {\bibinfo {title} {Determination of the proton spectral function of $^{12}$c from $(e,e^\prime p)$ data}} (\bibinfo {year} {2024}),\ \Eprint {https://arxiv.org/abs/2407.18226} {arXiv:2407.18226 [nucl-th]} \BibitemShut {NoStop}%
\bibitem [{\citenamefont {Ajimura}\ \emph {et~al.}(2020)\citenamefont {Ajimura} \emph {et~al.}}]{Ajimura:2020qni}%
  \BibitemOpen
  \bibfield  {author} {\bibinfo {author} {\bibfnamefont {S.}~\bibnamefont {Ajimura}} \emph {et~al.},\ }\bibfield  {title} {\bibinfo {title} {{Proposal: JSNS$^2$-II}},\ }\href@noop {} {\  (\bibinfo {year} {2020})},\ \Eprint {https://arxiv.org/abs/2012.10807} {arXiv:2012.10807 [hep-ex]} \BibitemShut {NoStop}%
\end{thebibliography}%
